%% file: main.tex
\documentclass{article}

\newcommand{\SysName}{Sylvie\xspace}
\newcommand{\SysNameP}{Sylvie-A\xspace}
\newcommand{\SysNameS}{Sylvie-S\xspace}

\input{math_commands.tex}

\input{packages}
\usepackage{hyperref}
\usepackage{url}

\usepackage{microtype}
\usepackage{graphicx}
\usepackage{subfigure}
\usepackage{booktabs} 

\usepackage{hyperref}

\usepackage[accepted]{mlsys2023}

\mlsystitlerunning{Boosting Distributed Full-Graph GNN Training with Asynchronous One-bit Communication}

\begin{document}

\twocolumn[
\mlsystitle{Boosting Distributed Full-Graph GNN Training with Asynchronous One-bit Communication}



\mlsyssetsymbol{equal}{*}

\begin{mlsysauthorlist}
\mlsysauthor{Meng Zhang}{ntu}
\mlsysauthor{Qinghao Hu}{ntu,sh}
\mlsysauthor{Peng Sun}{sh,sensetime}
\mlsysauthor{Yonggang Wen}{ntu}
\mlsysauthor{Tianwei Zhang}{ntu}
\end{mlsysauthorlist}

\mlsysaffiliation{ntu}{Nanyang Technological University}
\mlsysaffiliation{sensetime}{SenseTime Research}
\mlsysaffiliation{sh}{Shanghai AI Laboratory}

\mlsyscorrespondingauthor{Tianwei Zhang}{tianwei.zhang@ntu.edu.sg}

\mlsyskeywords{Machine Learning, MLSys}

\vskip 0.3in

\input{0_Abstract}

]
\printAffiliationsAndNotice{} 




\input{1_Introduction}
\input{2_Background_and_Motivation}

\input{3_System_Design}
\input{4_Evaluation}
\input{5_Discussion}
\input{7_Conclusion}

\bibliography{ref}
\bibliographystyle{mlsys2023}

\appendix
\newpage


\end{document}

%% file: math_commands.tex
\usepackage{amsmath,amsfonts,bm}

\def\eqref#1{equation~\ref{#1}}

\def\1{\bm{1}}

\DeclareMathAlphabet{\mathsfit}{\encodingdefault}{\sfdefault}{m}{sl}
\SetMathAlphabet{\mathsfit}{bold}{\encodingdefault}{\sfdefault}{bx}{n}



%% file: packages.tex
\usepackage{xspace}
\usepackage{mathtools} 
\usepackage{amssymb}
\usepackage{bm}
\usepackage{enumitem}
\usepackage{cite}
\usepackage{textcomp}
\usepackage{xcolor}
\usepackage{changepage}

\usepackage{graphicx}
\usepackage{tikz}

\usepackage{tabularx}
\usepackage{multirow}
\usepackage{booktabs}
\usepackage{makecell}
\usepackage[figuresleft]{rotating}
\usepackage[flushleft]{threeparttable}

\usepackage[utf8]{inputenc} 
\usepackage{tcolorbox} 
\usepackage{soul} 
\usepackage{pifont} 
\usepackage{bbding} 

\newcommand{\onex}{\raisebox{-0.6mm}{\large{\ding{182}}}}
\newcommand{\twox}{\raisebox{-0.6mm}{\large{\ding{183}}}}
\newcommand{\threex}{\raisebox{-0.6mm}{\large{\ding{184}}}}
\newcommand{\fourx}{\raisebox{-0.6mm}{\large{\ding{185}}}}
\newcommand{\fivex}{\raisebox{-0.6mm}{\large{\ding{186}}}}

\usepackage[noindentafter]{titlesec}
\titlespacing\section{0pt}{3pt}{3pt} 
\titlespacing\subsection{0pt}{3pt}{3pt}
\titlespacing\subsubsection{0pt}{3pt}{3pt}

\usepackage[skip=1pt]{caption} 

%% file: 0_Abstract.tex
\begin{abstract}

Training Graph Neural Networks (GNNs) on large graphs is challenging due to the conflict between the high memory demand and limited GPU memory. Recently, distributed full-graph GNN training has been widely adopted to tackle this problem. However, the substantial inter-GPU communication overhead can cause severe throughput degradation. Existing communication compression techniques mainly focus on traditional DNN training, whose bottleneck lies in synchronizing gradients and parameters. We find they do not work well in distributed GNN training as the barrier is the layer-wise communication of features during the forward pass \& feature gradients during the backward pass. To this end, we propose an efficient distributed GNN training framework \textbf{\SysName}, which employs one-bit quantization technique in GNNs and further pipelines the curtailed communication with computation to enormously shrink the overhead while maintaining the model quality. In detail, \SysName provides a lightweight \textit{Low-bit Module} to quantize the sent data and dequantize the received data back to full precision values in each layer. Additionally, we propose a \textit{Bounded Staleness Adaptor} to control the introduced staleness to achieve further performance enhancement. We conduct theoretical convergence analysis and extensive experiments on various models \& datasets to demonstrate \SysName can considerably boost the training throughput by up to 28.1$\times$.



\end{abstract}

%% file: 1_Introduction.tex
\section{Introduction}
\label{sec_intro}

In recent years, GNNs have become very popular and showed state-of-the-art (SOTA) performance in learning structured data like graphs. GNNs capture the underlying dependencies of the given graph via message passing operations \cite{GNN-survey}. To update nodes, GNNs first aggregate the feature vectors from the nodes' neighbors and then combine them together. Despite their impressive performance on graph-related tasks, training GNNs on large-scale graphs containing millions to billions of nodes is still a long-standing issue, as extensive memory resources are needed for loading and computing input graphs \cite{Link18,EXACT,GCN}, hindering the practical development of more sophisticated GNN models. 

Existing solutions to this problem can be divided into two directions. First, some works \cite{ClusterGCN,GraphSAINT,Layerdependent19,FastGCN,Adaptive18} propose sampling-based methods which only select a subset of nodes and edges to be trained at each iteration. However, most of these methods need centralized data storage, which will cause significant data transfer costs. More importantly, these methods suffer from model accuracy loss \cite{GraphSAGE,ROC,BNS-GCN}. Table \ref{table_sampling_acc} shows the test accuracy of the GraphSAGE model on the Ogbn-products dataset \cite{OGB} when training with the sampling-based mode and full-graph. The accuracy of sampling-based mode is always lower than that of full-graph training, especially as the sample size decreases.

\begin{figure}[t]
    \centering
    \includegraphics[width=\linewidth]{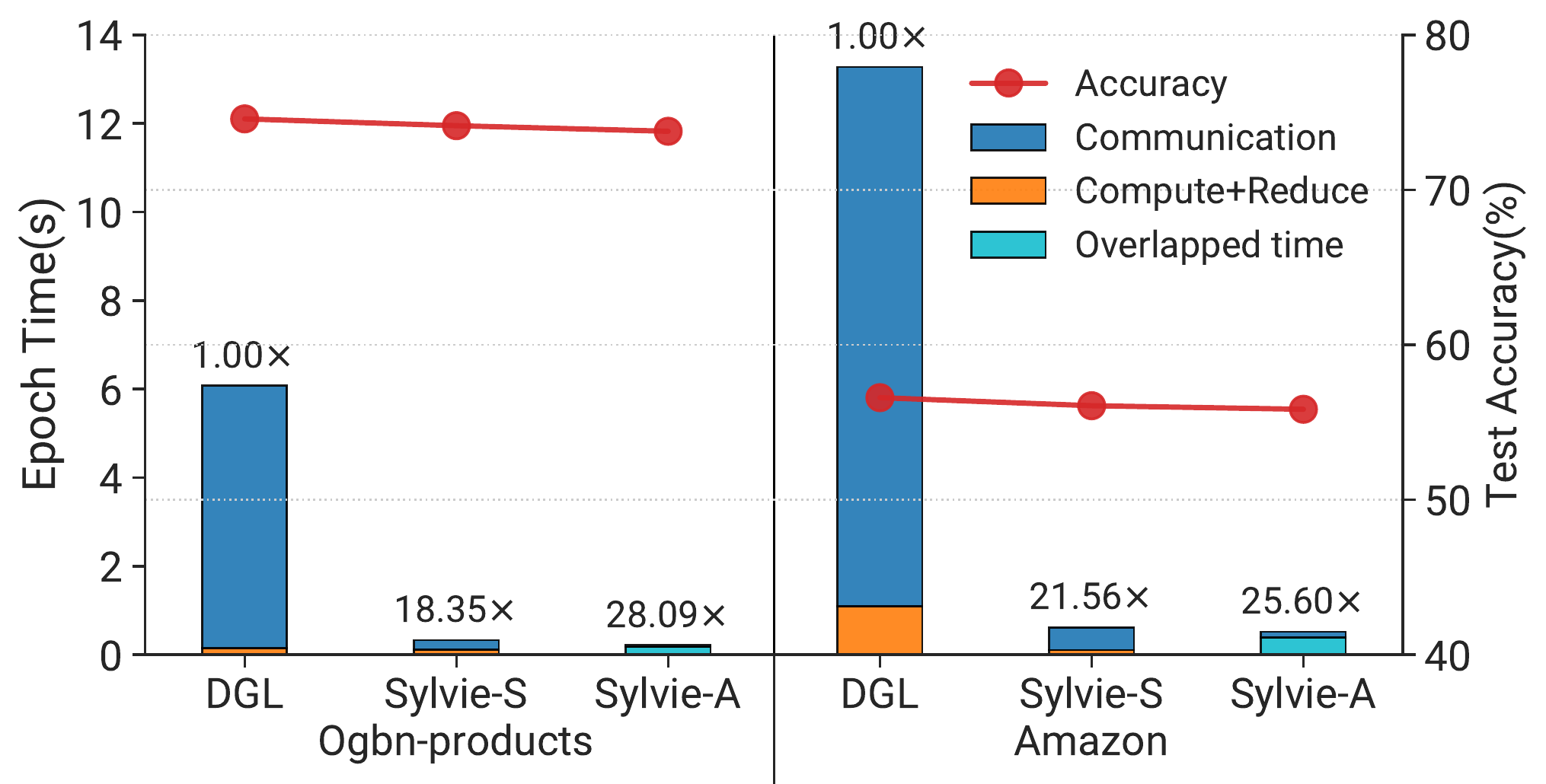}
    \vskip -10pt
    \caption{Epoch time and test accuracy of training GCN on two datasets with DGL, \SysNameS and \SysNameP over two servers. \SysName provides up to 28.1$\times$ speedup.}
    \label{fig_firstintro}
\end{figure}

\begin{table}[t]
    \centering
    \caption{Test accuracy of full-graph training and sampling-based training with different numbers of neighborhood samples in each layer when training GraphSAGE on Ogbn-products.}
    \vskip 0.1in
    \resizebox{0.7\linewidth}{!}{
      \begin{tabular}{@{}ccccc@{}}
\toprule
                       & \multicolumn{3}{c}{\textbf{Sampling-based}} & \multirow{2}{*}{\textbf{Full-Graph}} \\ \cmidrule(lr){2-4}
                     Sample Size  & 5           & 10          & 15         &                                      \\ \midrule
\textbf{Accuracy (\%)} & 73.55       & 74.87       & 76.84      & \textbf{79.19}                                \\ \bottomrule
\end{tabular}}
\label{table_sampling_acc}
\end{table}

Distributed GNN training \cite{Dorylus,DistDGL,AliGraph} is another promising direction, which conducts full graph training on multiple GPUs or nodes to reduce the computing time and memory demand on each GPU. It first splits the whole graph into several subgraphs so that each can fit in a single GPU, and then trains subgraphs on each GPU locally. Current GNN frameworks like DGL \cite{DGL} and PyG \cite{PyG} have the scalability limitation and lack system support for distributed full-graph training. In addition, unlike classical distributed DNN training (e.g., image classification with AlexNet \cite{TernGrad}) where training samples are independent of each other, it is non-trivial to apply data parallelism on GNNs due to the node dependency between subgraphs, leading to obligatory data communication overhead. To solve this, recent works \cite{BNS-GCN,CAGNET,ROC,SAR,PipeGCN} proposed various approaches to realize efficient distributed full-graph training. However, these approaches also exhibit several drawbacks. Prior methods \cite{CAGNET,ROC,SAR} still suffer from heavy communication overhead, resulting in considerable training time. Some works \cite{BNS-GCN,SAR} train large-scale graphs at the price of potential accuracy loss. Moreover, it is hard to generalize some approaches \cite{CAGNET,BNS-GCN,PipeGCN} to various GNN models or datasets.
To cope with the aforementioned limitations, we propose a new framework \SysName to effectively accelerate distributed full-graph GNN training while maintaining the model accuracy. 
\SysName consists of five main modules: (1) \textit{Graph Engine} first partitions the original graph to GPUs and selects the communicated nodes. (2) \textit{Quantizer} compresses the communicated features and gradients. (3) \textit{Communicator} transfers data between partitions. (4) \textit{Dequantizer} decompresses the received data in an error-compensated way. (5) \textit{Trainer} finally conducts model forward and backward computation using the communicated data. We also introduce two variants of \SysName: (1) \SysNameS executes the above modules sequentially. (2) \SysNameP integrates the asynchronous pipeline technique to further improve the performance. We also design a \textit{Bounded Staleness Adaptor} for \SysNameP to enhance the model convergence in some cases where convergence is affected by stale features \& gradients. Fig.\ref{fig_firstintro} shows part of the results when training with \SysName on two machines. \SysName dramatically reduces the training time with minor accuracy loss. The extensive communication overhead (blue bar) is cut down to a very tiny portion, thus contributing to the significant speedup on the epoch time. On top of \SysNameS, the curtailed communication is further overlapped by \SysNameP, achieving up to 28.1$\times$ speedup. In addition to the scalability over the multi-server setting, \SysName also achieves great performance on a single machine: e.g., up to 9.3$\times$ improvement over DGL.

In summary, we make the following contributions:
\begin{itemize}[leftmargin=*, itemsep=0pt, topsep=0pt]
    \item To the best of our knowledge, \SysName is the \textit{first} framework to implement one-bitwidth communication quantization to accelerate multi-GPU GNN training, achieving up to 28.1$\times$ speedup over existing GNN frameworks.

    \item Based on our plentiful experiments and theoretical analysis, \SysName can well maintain the model convergence with negligible accuracy loss.

    \item Beyond the \emph{spacial} size reduction, we also consider the orthogonal \emph{temporal} aspect by integrating the asynchronous pipeline technique (\SysNameP). Additionally, for better model convergence, we design a \emph{Bounded Staleness Adaptor} to reduce the errors incurred by stale data. 
\end{itemize}

%% file: 2_Background_and_Motivation.tex
\section{Background and Related Work}
\label{sec_motivation}

\subsection{Graph Neural Networks}
GNNs are machine learning algorithms that learn from the graph connectivity and model the relationship between nodes. In general, the iterative learning process contains two steps in each layer: feature aggregation and update. Consider a graph $\mathcal{G}=(V, E)$ with nodes $V=\left\{v_1, \cdots, v_{|V|}\right\}$, edges $E=\left\{e_1, \cdots, e_{|E|}\right\}$ and node feature matrix $\boldsymbol{X} \in \mathbb{R}^{|V| \times d}$. For an arbitrary layer $l \in [1, L]$, the aggregation and update steps can be expressed as:
\begin{align}
    \boldsymbol{z}_v^{(l)}&=\rho^{(l)}\left(\left\{\boldsymbol{h}_u^{(l-1)} \mid u \in \mathcal{N}(v)\right\}\right)\label{aggre} \\
    \boldsymbol{h}_v^{(l)}&=\phi^{(l)}\left(\boldsymbol{z}_v^{(l)}, \boldsymbol{h}_v^{(l-1)}\right)\label{update}
\end{align}
where $\mathcal{N}(v)$ is the neighboring nodes of node $v$. The aggregation function $\rho^{(l)}$ takes the embeddings of neighboring nodes $\boldsymbol{h}_u^{(l-1)}$ to get an intermediate aggregated result $\boldsymbol{z}_v^{(l)}$, which then serves as the input to update function $\phi^{(l)}$ together with the feature vector $\boldsymbol{h}_v^{(l-1)}$ of node $v$ itself to obtain the learned embedding $\boldsymbol{h}_v^{(l)}$ at the $l$-th layer. Different GNNs vary in their aggregation and update functions. In this work, we mainly focus on three popular GNN models: GraphSAGE \cite{GraphSAGE}, GCN \cite{GCN} and GAT \cite{GAT}, but our framework can be easily extended to other GNN models. 

\subsection{Distributed GNN Training}
\label{sec_distributedGNN}
\begin{figure}[t]
    \centering
    \includegraphics[width=0.5\textwidth]{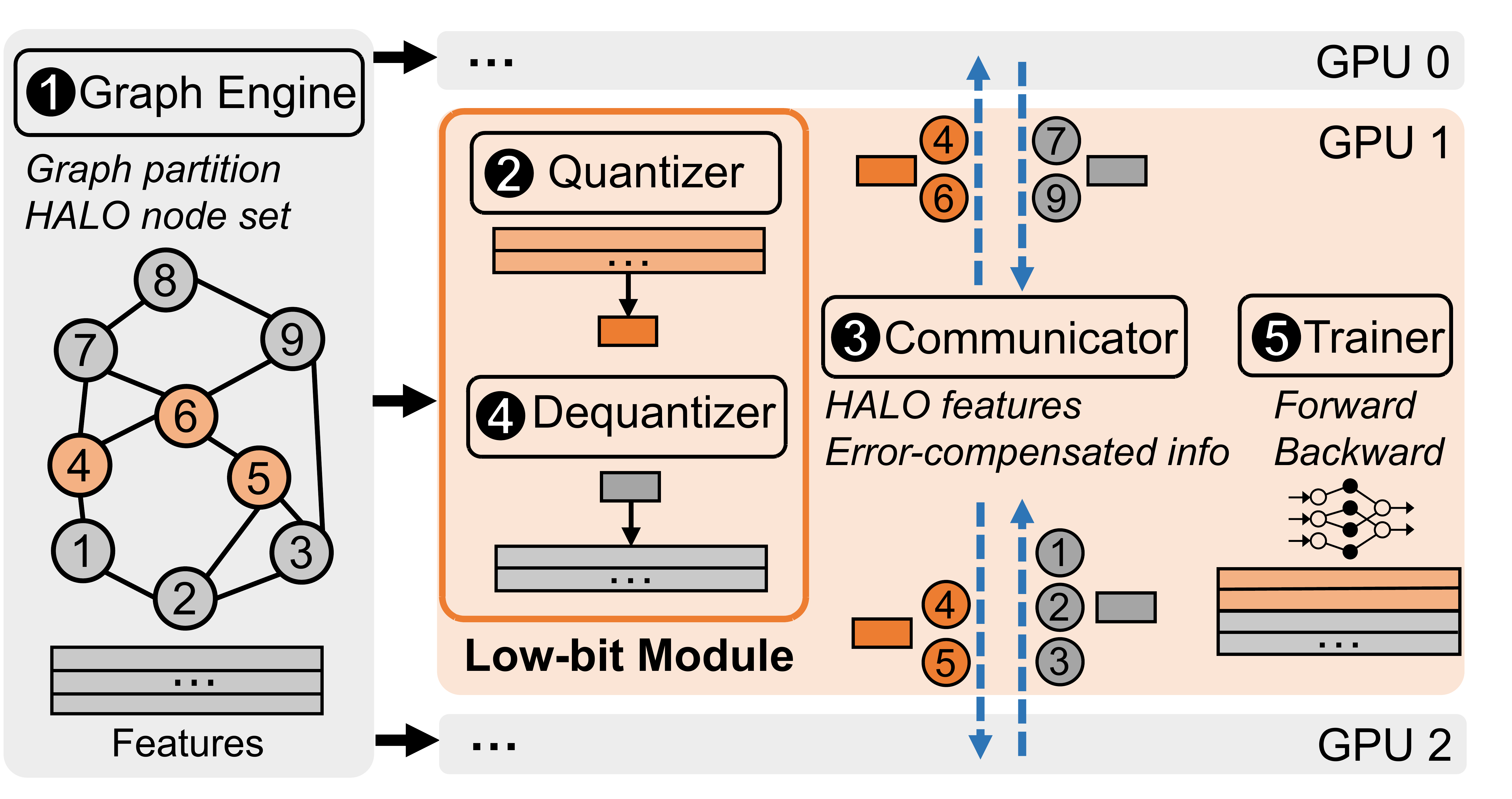}
    \vskip -10pt
    \caption{Overview of \SysName framework \& workflow. We focus on the subgraph on GPU-1. Orange nodes and rectangles represent nodes allocated to GPU-1 and their corresponding features. The others in gray represent nodes/features on other GPUs. }
    \label{1BG}
\end{figure}

To conduct distributed GNN training on full graphs, the whole input graph is first partitioned on the host side to fit into a single GPU. Since each node and its features will only be assigned to one GPU, there are some nodes, dubbed \textbf{HALO nodes}, that are connected to nodes in the local partition but do not belong to this partition. As depicted by Fig.\ref{1BG}, for the partition on GPU-1, node 4 requires extra features of node 7 from GPU-0 and node 1 from GPU-2 to update its embedding in every layer. In the backward pass, the feature gradients computed locally are also incomplete. Since nodes only exchange their 1-hop neighbors in each layer, the dataflow from 2-hop to $L$-hop resides on other partitions. Therefore, the feature gradients of HALO nodes will also be broadcast in each layer. This communication overhead is non-trivial since the amount of HALO nodes can be excessive. In addition, such cost becomes more intensive as the number of partitions and layer size grow larger. For node 4 on GPU1, communication of nodes 7 and 1 should be done before layer1 begins. After layer1, the same process repeats for the rest layers in a sequential order strictly, as shown in Fig.\ref{workflow}(a). In this case, excessive communication will take up the training time and block the subsequent computation.

\begin{figure}[t]
    \centering
    \includegraphics[width=0.4\textwidth]{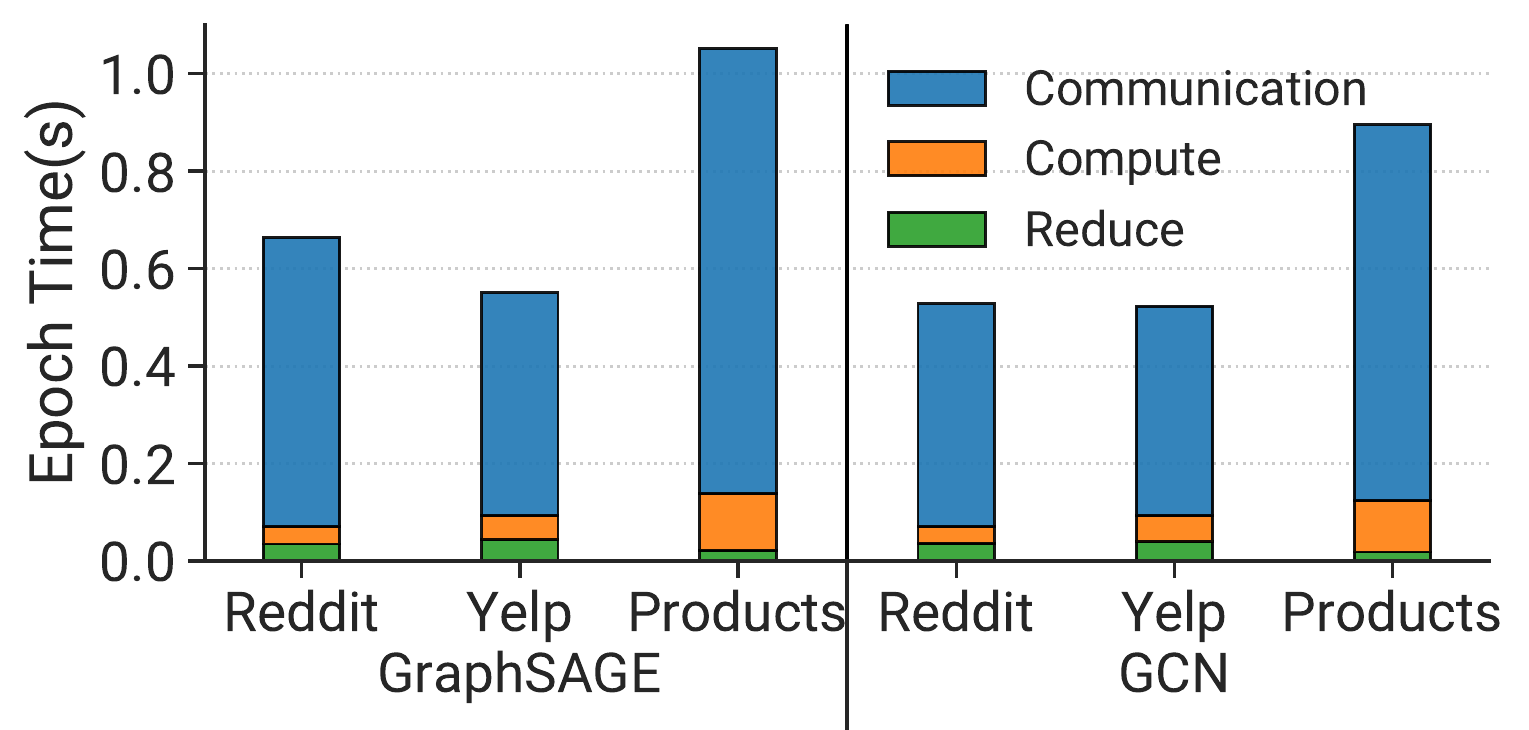}
    \vskip -10pt
    \caption{Training time per epoch and its breakdown in vanilla distributed GNN training with DGL on a single server (8 GPUs).}
    \label{fig_breakdown}  
\end{figure}

To show the vast communication cost more intuitively, we profile the epoch time along with its breakdown on multiple datasets and models, as shown in Fig.\ref{fig_breakdown}. We can clearly see for both GraphSAGE and GCN on three datasets, the communication time nearly dominates the entire training process (up to 89.23\%), while the computation and gradient synchronization operation (all-reduce) only occupy a very small portion. The scalability and efficiency of distributed GNN training is thus seriously restrained due to the excessive communication overhead.

Prior works propose new frameworks to accelerate distributed GNN training, e.g., AliGraph \cite{AliGraph} and NeuGraph \cite{NeuGraph}. However, these methods all store the partitions in CPUs, which inevitably incur frequent CPU-GPU swapping and largely impair the benefits of distributed training. DistDGL \cite{DistDGL} provides the scaling results but only on sampling-based methods. LLCG \cite{LLCG} totally drops dependent information between partitions and adds a global correction server to compensate for the error with redundant overhead. Moreover, those works only support mini-batch training on graphs rather than full-graph training.

Different from the above sampling-based works, ROC \cite{ROC} accelerates distributed full-graph training, but it also stores the partitions in CPUs and suffers from the expensive CPU-GPU data transfer cost. SAR \cite{SAR} provides memory savings for full-graph training in large scale, but it has more computation burden due to rematerialization. BNS-GCN \cite{BNS-GCN} adopts random sampling on the boundary nodes and shows impressive acceleration, yet it risks downgrading the model performance by dropping node connections and its performance is highly dependent on the graph structure. CAGNET \cite{CAGNET} raises different dimension partitioning methods to boost training by slicing the node features to sub-vectors, at the extra communication and synchronization costs. PipeGCN \cite{PipeGCN} ingeniously hides the communication cost by pipelining computation and communication while keeping all the boundary information. However, its efficacy corrupts badly when the communication time is far larger than the computation part, as illustrated in \S \ref{eval_compare}.

\subsection{Quantization for Neural Networks}
Model quantization is commonly adopted to accelerate conventional DNN inference \cite{QuantizeDNN,QuantizeInfer}, while we target training speedup. Compression has also been popular to reduce the communication and boost distributed DNN training \cite{TernGrad,QSGD}. However, the overhead in distributed DNN training stems from synchronizing all gradients and parameters. Hence, they cannot be simply grafted to our scenario since the overhead of gradient synchronization like all-reduce is negligible in GNN training, as shown in Fig.\ref{fig_breakdown}.

In recent years, some works \cite{SGQuant,LowGNN} apply GNN model quantization via simulation for memory reduction, with the underlying computation still in 32-bit full precision. A recent work EXACT \cite{EXACT} aims to reduce the memory demand at the cost of extra training time overhead, seriously deteriorating the training efficiency. Other works \cite{Degree-Quant,QGTC} quantize GNN models for efficient inference. Compared with our work, all these methods have different targets and only consider small-scale datasets. 


Different from the aforementioned quantization works, we are the \textbf{first} to explore the opportunity of quantizing communication in GNNs, which can substantially reduce the overhead with negligible accuracy loss. We do not quantize activations or weights like previous works \cite{QuantizeDNN,QuantizeInfer} because (1) computation only occupies a very small portion while communication of embeddings \& feature gradients dominates the training overhead (Fig.\ref{fig_breakdown}); (2) unique sparse computations in GNNs, e.g., SpMM in cuSPARSE \cite{cuSPARSE}, lack support for low precision computation, unlike dense operations (e.g. GEMM) in DNNs which support fast low precision computation. 

%% file: 3_System_Design.tex
\section{Proposed Framework}
\label{sec_design}

\subsection{Overview}
Fig.\ref{1BG} depicts the overall workflow of \SysName, consisting of five steps. \onex{} \textit{Graph Engine} first partitions the input graph to several subgraphs on the host side, and constructs the HALO node set of each partition for the preparation of later communication. Then each partition, including the graph structure and features, is allocated to an individual GPU. \twox{} Next, \textit{Quantizer} in each GPU quantizes the HALO node data into 1-bit integers. For instance, data of nodes 4, 5 \& 6 in GPU-1 are quantized because they are required by other partitions in GPU-0 and GPU-1. \threex{} After this, those quantized data, along with the corresponding error-compensated information (used to help recover data), are broadcast between partitions through network communications. \fourx{} After receiving the quantized data from other partitions, \textit{Dquantizer} in each GPU recovers the compressed data back to full-precision values with the error-compensated information. \textit{Quantizer} and \textit{Dequantizer} together form the \textit{Low-bit Module} that takes charge of data transformation. \fivex{} Those recovered data, together with the original ones are utilized to train the model in both forward and backward passes.

\begin{figure}[t]
    \centering
    \includegraphics[width=0.5\textwidth]{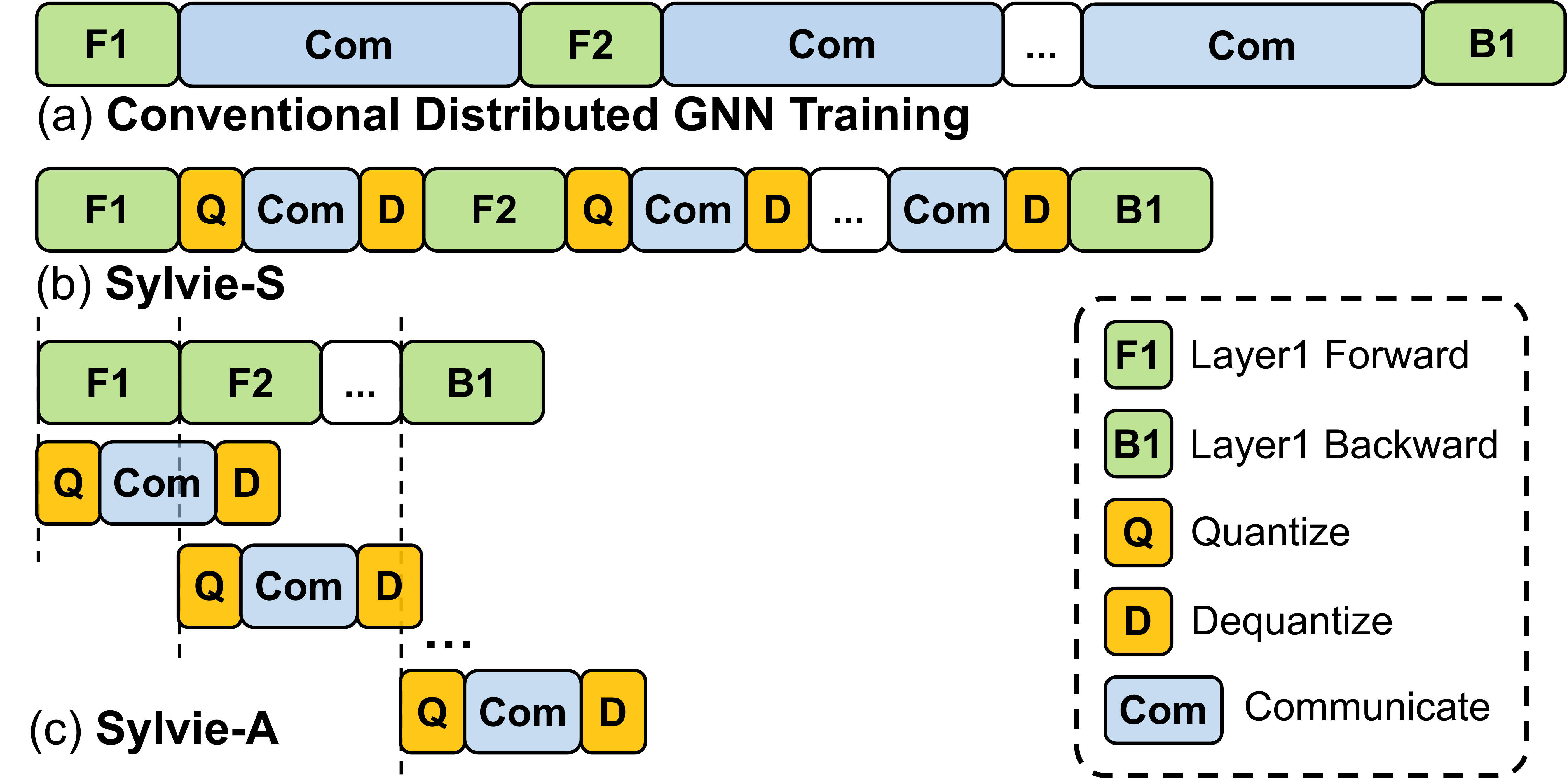}
    \vskip -10pt
    \caption{Workflow comparison between conventional distributed training, \SysNameS and \SysNameP. \SysName reduces the massive communication overhead with \textit{Low-bit Module}. \SysNameP further curtails the training time by pipelining.}
    \label{workflow}
\end{figure}

We introduce two variants based on \SysName (Fig.\ref{workflow}). (1) \textbf{\SysNameS}: all these modules are executed synchronously in a sequential order. (2) \textbf{\SysNameP}: we adopt the asynchronous pipeline technique between \textit{Low-bit Module}, communication and computation to further improve the training performance. However, this asynchronous pipeline can cause damages to the convergence on some datasets or models, due to the stale features and gradients. We further design a \textit{Bounded Staleness Adaptor} to  periodically synchronize the latest data for all partitions in a fixed number of epochs.

\subsection{\SysName Framework}

\textbf{Graph Engine}. To conduct distributed GNN training, \textit{Graph Engine} first partitions both the graph structure and feature matrix on the host side based on the number of available GPUs. The HALO node set of each partition is also constructed. Then each partition is loaded to an individual GPU, and each GPU holds a replica of the full model.

\textbf{Quantizer}. Different from existing works that quantize all the activations or weights, \SysName is the first to quantize the communication data in distributed GNN training. We illustrate the quantization mechanism using the example of the subgraph on GPU-1 during the forward pass in Fig.\ref{1BG}. For the $l$-th layer of the model, \textit{Quantizer} first selects node features needed for communication (e.g., features of nodes 4, 5 and 6) $\boldsymbol{S}^{(l-1)} \subseteq \boldsymbol{H}^{(l-1)}$ according to the HALO node set and then quantizes them to 1-bit integers $\hat{\boldsymbol{S}}^{(l-1)}_{1bit}$ with low precision. For each node feature $\boldsymbol{h}^{(l-1)}$ in $\boldsymbol{S}^{(l-1)}$, to quantize it to a 1-bit integer, we use the following formula:
\begin{align}
    \bar{\boldsymbol{h}}^{(l-1)}&=\frac{\boldsymbol{h}^{(l-1)}-\min(\boldsymbol{h}^{(l-1)})}{\max(\boldsymbol{h}^{(l-1)})-\min(\boldsymbol{h}^{(l-1)})} B \label{quantize} \\
    \hat{\boldsymbol{h}}_{1bit}^{(l-1)}&=\left\lceil \bar{\boldsymbol{h}}^{(l-1)} \right\rceil \quad \mathrm{w.prob.} \quad \bar{\boldsymbol{h}}^{(l-1)}- \left\lfloor \bar{\boldsymbol{h}}^{(l-1)} \right\rfloor \notag \\
    & \qquad \qquad \qquad \mathrm{otherwise}\quad \left\lfloor \bar{\boldsymbol{h}}^{(l-1)} \right\rfloor \label{rounding}
\end{align}
where $B=2^b-1$ is the number of quantization bins when we quantize data to $b$-bit integers, and the default value of $b$ is 1 in our work since we make 1-bit quantization. Equ.\ref{rounding} is the stochastic rounding \cite{BinaryConnect}. 

\textbf{Communicator}. After the quantization, $\hat{\boldsymbol{S}}^{(l-1)}_{1bit}$, which contains all quantized $\hat{\boldsymbol{h}}_{1bit}^{(l-1)}$ is broadcast to the corresponding GPU via \textit{Communicator}. Besides, \textit{Communicator} also sends error-compensated information such as scale (the ratio between $\max(\boldsymbol{h}^{(l-1)})-\min(\boldsymbol{h}^{(l-1)})$ and $B$) and zero-point $\min(\boldsymbol{h}^{(l-1)})$ along with the quantized data, to better recover data unbiasedly. As shown in Table \ref{table_commu_volume}, the error-compensated data have much smaller size compared with the main data of communicated embeddings or gradients, incurring negligible communication overhead.

\textbf{Dequantizer}. Managed by \textit{Communicator}, each partition itself receives the compressed features $\hat{\boldsymbol{R}}^{(l-1)}_{1bit}$ along with error-compensated information from other partitions. Then \textit{Dequantizer} dequantizes them back to 32-bit full precision floating-point values $\tilde{\boldsymbol{R}}^{(l-1)}$, with the help of error-compensated information, to ensure the consistency between the recovered features and the original ones as much as possible. The dequantization process of each embedding $\hat{\boldsymbol{h}}_{1bit}^{(l-1)}$ in $\hat{\boldsymbol{R}}^{(l-1)}_{1bit}$ is expressed as:
\begin{equation}
\tilde{\boldsymbol{h}}^{(l-1)}=\frac{(\max(\boldsymbol{h}^{(l-1)})-\min(\boldsymbol{h}^{(l-1)}))\hat{\boldsymbol{h}}_{1bit}^{(l-1)}}{B}+\min(\boldsymbol{h}^{(l-1)}) \label{dequantize}
\end{equation}
where the recovered $\tilde{\boldsymbol{R}}^{(l-1)}$ includes all dequantized $\tilde{\boldsymbol{h}}^{(l-1)}$. 

\textbf{Trainer}. Using $\tilde{\boldsymbol{R}}^{(l-1)}$ together with $\boldsymbol{H}^{(l-1)}$ which is already stored in this partition, node embeddings $\boldsymbol{H}^{(l)}$ of $l$-th layer are updated according to Equ.\ref{aggre} and \ref{update} by \textit{Trainer}. 

The backward pass shares a similar training process as above, except that the communicated data are feature gradients instead, and \textit{Trainer} gradually computes the gradients after the received data are dequantized. Fig.\ref{workflow}(b) depicts the workflow of \SysNameS in each partition as described above. It vastly cuts down the communication overhead compared with the conventional distributed training shown in Fig.\ref{workflow}(a). For \SysNameS, computation in each layer follows the communication step, and the same process occurs on other partitions. 

\subsection{Asynchronous Curtailed Communication with Computation}
To further reduce the epoch training time and cut down the overhead brought by quantization and dequantization operations, inspired by \citet{PipeGCN}, we propose \SysNameP by pipelining the \textit{Low-bit Module} and communication with GNN computation on top of \SysName. Fig.\ref{workflow} (c) depicts the detailed workflow of \SysNameP. Different from the vanilla distributed training and \SysNameS, \SysNameP directly begins each layer's computation with existing information in this partition. Meanwhile, \textit{Low-bit Module} concurrently quantizes data and communicates the compressed data. Thus, \SysNameP breaks the sequential order between computation and communication and realizes distributed GNN training in an asynchronous way.
Note that in \SysNameP, the currently overlapped communication will be used in the next epoch, ensuring the data integrity when the computation starts. 

\textbf{Bounded Staleness Adaptor}. Asynchronization could introduce stale embeddings and feature gradients, which may affect the model convergence. Although the model convergence is robust against staleness for most datasets or models as shown in \S \ref{eval_converge}, it is still important to mitigate the negative effects on the convergence in other cases. \SysNameP implements a simple yet effective \textit{Bounded Staleness Adaptor} to perform compulsory synchronization of embeddings or gradients periodically to control the errors brought by staleness. Though the bounded staleness has already been investigated in traditional distributed ML works \cite{H-PS} to improve SGD convergence, the measures to limit staleness in distributed GNN training have been rare. Specifically, this module conducts one sequential training after a fixed epoch interval $\epsilon_s$ to synchronize the latest embeddings or gradients, where $\epsilon_s$ can be defined by users as a trade-off between the training throughput and convergence rate. For the rest epochs, the training is still in the pipeline mode. \SysNameP with \textit{Bounded Staleness Adaptor} enabled are denoted as \SysName-A$\epsilon_s$. Experiment results in Fig.\ref{fig_stale_adaptor} demonstrate that \textit{Bounded Staleness Adaptor} well improves the convergence rate when using the asynchronous pipeline technique.

\subsection{GNN Training with \SysName}
\label{algo}
We elaborate how \SysNameS works on GNNs in the forward and backward passes. 

\textbf{Forward Propogation}. The forward pass of \SysNameS in each partition is outlined in Alg.\ref{algo_forward}. The initial node embeddings are set as the feature vectors (line 2). First, we obtain HALO node set so that nodes will be delivered correctly later. Within $T$ epochs, in each GNN layer, a vertex updates its embedding with its one-hop neighbors' embeddings of the last layer, which may include HALO nodes whose embeddings reside in different partitions. Those embeddings are quantized by \textit{Quantizer} in their own partitions (line 10), distributed through network communications (line 11), and then dequantized by \textit{Dequantizer} in the current partition (line 12). The recovered embeddings, together with those originally stored in this GPU, are concatenated into $\tilde{\boldsymbol{H}}^{(l-1)}_n$ to update the embeddings. Line 15 in Alg.\ref{algo_forward} is the matrix format of Equ.\ref{aggre} and \ref{update}, where $\boldsymbol{A}_n=\boldsymbol{D}^{-\frac{1}{2}}(\boldsymbol{A}_n+\boldsymbol{I}) \boldsymbol{D}^{-\frac{1}{2}}$ is the normalized adjacency matrix, $\boldsymbol{D}$ is the degree matrix, and $\boldsymbol{I}$ is the identity matrix. 

\begin{algorithm}[tb]
    \caption{Forward phase of \SysNameS}
    \small
    \label{algo_forward} 
    \begin{algorithmic}[1]
        \INPUT Partition id $n$, subgraph $\mathcal{G}_n$, feature matrix $\boldsymbol{X}_n$, label $Y_n$, adjacency matrix $\boldsymbol{A}_n$, epoch number $T$, layer size $L$, local node set $V_n$, weights $\boldsymbol{W}^{(l-1)}$ 
        \STATE \textbf{Partition} $n=1,2,...,N$ in Parallel: 
        \STATE $\boldsymbol{H}^{(0)}_n=\boldsymbol{X}_n$ \COMMENT{Initialize node embeddings} \label{line2}
        \STATE $V_{\mathrm{HALO},n}=\{$node $v \in \mathcal{G}_n:v \notin V_n\}$ 
        \STATE  Distribute $V_{\mathrm{HALO},n}$ and receive $[V_{\mathrm{HALO},1}, ..., V_{\mathrm{HALO},N}]$
        \STATE  Distribute $V_n$ and receive $[V_{1}, ..., V_{N}]$
        \STATE  $[S_1,...,S_N]=[V_n\cap V_{\mathrm{HALO},1}, ..., V_n\cap V_{\mathrm{HALO},N}]$ 
        \STATE  $[R_1,...,R_N]=[V_{\mathrm{HALO},n}\cap V_1,...,V_{\mathrm{HALO},n}\cap V_N]$ 

        \FOR{$t$ \textbf{from} $1$ \textbf{to} $T$ }
        \FOR{$l$ \textbf{from} $1$ \textbf{to} $L$ }
        \STATE $\hat{\boldsymbol{S}}^{(l-1)}_{1bit}=$ quantize$([\boldsymbol{H}^{(l-1)}_n(S_1),...,\boldsymbol{H}^{(l-1)}_n(S_N)])$ 
        \STATE Send $\hat{\boldsymbol{S}}^{(l-1)}_{1bit}$ to partition $1,...,N$. Receive $\hat{\boldsymbol{R}}^{(l-1)}_{1bit}$ from partition $1,...,N$
        \STATE $\tilde{\boldsymbol{R}}^{(l-1)}=$ dequantize$(\hat{\boldsymbol{R}}^{(l-1)}_{1bit})$ 
        \STATE $\tilde{\boldsymbol{H}}^{(l-1)}_n=$ concatenate$(\boldsymbol{H}^{(l-1)}_n, \tilde{\boldsymbol{R}}^{(l-1)})$
        \IF {$l \neq L$}
        \STATE $\boldsymbol{H}^{(l)}_n=\sigma\left(\boldsymbol{A}^\top_n \tilde{\boldsymbol{H}}^{(l-1)}_n \boldsymbol{W}^{(l)}_{t-1}\right)$ \COMMENT{Update embeddings}
        \ELSE 
        \STATE $\boldsymbol{H}^{(l)}_n=\boldsymbol{A}^\top_n \tilde{\boldsymbol{H}}^{(l-1)}_n \boldsymbol{W}^{(l)}_{t-1}$

        \ENDIF
        \ENDFOR
        \STATE $\mathcal{L}=$Loss(softmax$(\boldsymbol{H}^{(L)}_n), Y_n)$

        \ENDFOR
        
    \end{algorithmic}
\end{algorithm}

\textbf{Backward Propogation}. Alg.\ref{bp} shows the backward pass of \SysNameS. In the forward phase, the feature embeddings used for update contain those already stored and communicated ones from other partitions. Therefore, in the backward pass, the feature gradients computed locally is incomplete since the dataflow of more than 1-hop cannot be obtained locally. Therefore, from the $L$-th to the 1-st layer, the feature gradients $\boldsymbol{J}^{(l)}_n$ also need to be communicated between partitions. Similar as the forward phase, \textit{Quantizer} and \textit{Dequantizer} are in charge of quantizing and dequantizing feature gradients (lines 10 and 12). After integrating the gradients (line 13), the weights can be correctly computed (lines 17).

\begin{algorithm}[tb]
    \caption{Backward phase of \SysNameS}
    \small
    \label{bp} 
    \begin{algorithmic}[1]
        \INPUT Partition id $n$, label $Y_n$, adjacency matrix $\boldsymbol{A}_n$, epoch number $T$, layer size $L$, local node set $V_n$, weights $\boldsymbol{W}^{(l-1)}$
        \STATE \textbf{Partition} $n=1,2,...,N$ in Parallel:
        \FOR{$t$ \textbf{from} $1$ \textbf{to} $T$ }
        \FOR{$l$ \textbf{from} $L$ \textbf{to} $1$ }
        \IF {$l=L$}
        \STATE $\boldsymbol{J}^{(L)}_n=\nabla_{\boldsymbol{H}^{(L)}_n} \mathcal{L}$
        \ENDIF

        \STATE $\boldsymbol{G}^{(l)}_n=\left[\boldsymbol{A}_n \tilde{\boldsymbol{H}}^{(l-1)}_n\right]^{\top}\left(\boldsymbol{J}^{(l)}_n \circ \sigma^{\prime}\left(\boldsymbol{A}_n \tilde{\boldsymbol{H}}^{(l-1)}_n \boldsymbol{W}^{(l)}_{t-1}\right)\right)$ \COMMENT{Compute weight gradients}

        \IF {$l>1$}
        \STATE $\boldsymbol{J}^{(l-1)}_n = \boldsymbol{A}_n^{\top}\left(\boldsymbol{J}^{(l)}_n \circ \sigma^{\prime}\left(\boldsymbol{A}_n \tilde{\boldsymbol{H}}^{(l-1)}_n \boldsymbol{W}^{(l)}_{t-1}\right)\right)\left[\boldsymbol{W}^{(l)}_{t-1}\right]^{\top}$ \COMMENT {Compute feature gradients}
        \STATE $\hat{\boldsymbol{S}}^{(l-1)}_{1bit}=$ quantize$([\boldsymbol{J}^{(l-1)}_n(R_1),...,\boldsymbol{J}^{(l-1)}_n(R_N)])$ 
        \STATE Send $\hat{\boldsymbol{S}}^{(l-1)}_{1bit}$ to partition $1,...,N$. Receive $\hat{\boldsymbol{R}}^{(l-1)}_{1bit}$ from partition $1,...,N$
        \STATE $\tilde{\boldsymbol{R}}^{(l-1)}=$ dequantize$(\hat{\boldsymbol{R}}^{(l-1)}_{1bit})$ 
        \STATE $\boldsymbol{J}^{(l-1)}_n=\boldsymbol{J}^{(l-1)}_n + \tilde{\boldsymbol{R}}^{(l-1)}$ \COMMENT{Integrate feature gradients}
        \ENDIF

        \ENDFOR
        \STATE $\boldsymbol{G}=$AllReduce$(\boldsymbol{G_n})$ \COMMENT{Gradient synchronization}
        \STATE $\boldsymbol{W}_t = \boldsymbol{W}_{t-1}-\eta \boldsymbol{G}$ \COMMENT{Update model}

        \ENDFOR
        
    \end{algorithmic}
\end{algorithm}

\subsection{Theoretical Analysis}
\label{sec_converge_analysis}

In this section, we study the convergence of \SysName. We theoretically analyze the unbiased characteristic of \textit{Low-bit Module} and the limited introduced noise, so the model accuracy and convergence on \SysName could be well maintained.

\textbf{Unbiased Low-bit Module}. The following theorems characterize the unbiased feature of quantization, and are referenced from ActNN \cite{ActNN}. 
\newtheorem{theorem}{Theorem}
\begin{theorem}
    \label{theorem1}
    (Unbiased embeddings) Assume that $\bar{\boldsymbol{h}}^{(l)}- \left\lfloor \bar{\boldsymbol{h}}^{(l)} \right\rfloor \sim \mathcal{U}(0,1)$, $D$ is the hidden size of GNN layers, the quantized and dequantized embeddings are unbiased.
    \begin{align}
    \mathbb{E}\left[\tilde{\boldsymbol{h}}^{(l)}\right]&=\mathbb{E}[\mathrm{Dequantize}(\mathrm{Quantize}(\boldsymbol{h}^{(l)}))]=\boldsymbol{h}^{(l)} \label{expect} \\ 
    \mathrm{Var}(\tilde{\boldsymbol{h}}^{(l)})&=\frac{D\left[ \max(\boldsymbol{h}^{(l)})-\min(\boldsymbol{h}^{(l)})\right]^2}{6 B^2} \label{variance}
    \end{align}
\end{theorem}

\begin{theorem}
    \label{theorem2}
    (Unbiased gradients) There exist random quantization strategies $\hat{\mathbf{C}}$ that give unbiased weight gradients. 
   \begin{align}
    \mathbb{E}\left[\hat{\boldsymbol{G}}^{(l)}\right]&=\boldsymbol{G}^{(l)} \label{grad_expect} \\
    \mathrm{Var}[\hat{\boldsymbol{G}}^{(l)}]&=\mathrm{Var}\left[\boldsymbol{G}^{(l)}\right]+ \notag \\
    \sum_{m=l}^L \mathbb{E}&\left[\mathrm{Var}\left[\mathbf{B}^{(l \sim m)}\left(\hat{\boldsymbol{J}}^{(m)}, \hat{\mathbf{C}}^{(m)}\right) \mid \hat{\boldsymbol{J}}^{(m)}\right]\right] \label{grad_variance}
\end{align}
\end{theorem}
where $\mathbf{B}^{(l)}$ is the backward function of the $l$-th layer.

The two theorems prove the quantization and dequantization operations are unbiased, so are the subsequently calculated weights and gradients. Besides, the noise introduced by these two operations in \SysName is limited, so the model quality can be well kept.  Equ.\ref{variance} and \ref{grad_variance} reveal that quantization brings some extra noise to the training data, and the noise is inversely proportional to the number of bits. Though the noise will be aggregated layer-by-layer and result in a paramount accuracy drop (usually $>$5\%) to typical CNNs \cite{ActNN}, we find its influence on GNNs is negligible. The layer size of GNNs (usually 2 to 4) is far smaller than that of CNNs, making GNNs more noise-tolerant than CNNs \cite{EXACT}. Therefore, quantization is suitable to be applied to facilitate distributed GNN training. Table \ref{table_multinode} also shows the effect of the introduced noise on accuracy can be neglected in practice.

\textbf{Model Convergence}. Given the unbiased gradients, we can establish the convergence of \SysName. Suppose we have the common SGD in the form of $\boldsymbol{W}_{t+1} = \boldsymbol{W}_{t}-\eta \hat{\boldsymbol{G}}$, starting from the initial weights $\boldsymbol{W}_{1}$. We make the following assumptions: 
\begin{itemize}[leftmargin=*, itemsep=0pt, topsep=0pt]
\item The loss $\mathcal{L}$ is continuous differentiable and $\nabla \mathcal{L}\left(\boldsymbol{W}\right)$ is $\beta$-Lipschitz continuous \cite{LipschitzContinuity}. 
\item $\mathcal{L}$ is bounded below by $\mathcal{L}_{inf}$. 
\item There exists $\sigma^2>0$, such that $\forall \boldsymbol{W}, \operatorname{Var}\left[\hat{\boldsymbol{G}} \right] \leq \sigma^2$, where for any vector $\mathbf{x}, \operatorname{Var}[\mathbf{x}]:=\mathbb{E}\|\mathbf{x}\|^2-\|\mathbb{E}[\mathbf{x}]\|^2$.
\end{itemize}
Then we can have the following convergence theorem, taken from Theorem 4.8 in \citet{bottou2018optimization}.
\begin{theorem}
    \label{theorem3}
    (Convergence) If $0<\eta \leq \frac{1}{\beta}$, for iteration $t$ in $\{1,...,T\}$, where $T$ is the maximum number of iterations, we have
    \begin{equation}
        \mathbb{E}\left\|\nabla \mathcal{L}\left(\boldsymbol{W}_t\right)\right\|^2 \leq \frac{2\left(\mathcal{L}\left(\boldsymbol{W}_1\right)-\mathcal{L}_{i n f}\right)}{\eta T}+\eta \beta \sigma^2 \label{equa_converge}
    \end{equation}
\end{theorem}

The first term of Equ.\ref{equa_converge} converges to zero as the number of iterations $T$ goes to infinity. Therefore, the algorithm converges to the neighborhood of a stationary point, where the radius is controlled by the gradient variance.

In addition, different from most compression methods \cite{EXACT,ActNN}, in each epoch, \SysName only applies quantization to a portion of embeddings and gradients (i.e. those needed for communication), allowing for unbiased gradients to flow through weights. This controls the amount of noise so that the extreme 1-bit compression can be applied with limited accuracy loss. Similar techniques can be found in existing works \cite{QuantNoise,SQ,Degree-Quant} which adopt the subset quantization.



%% file: 4_Evaluation.tex
\section{Experiments}
\label{eval}
\begin{figure*}[t]
    \centering
    \includegraphics[width=\textwidth]{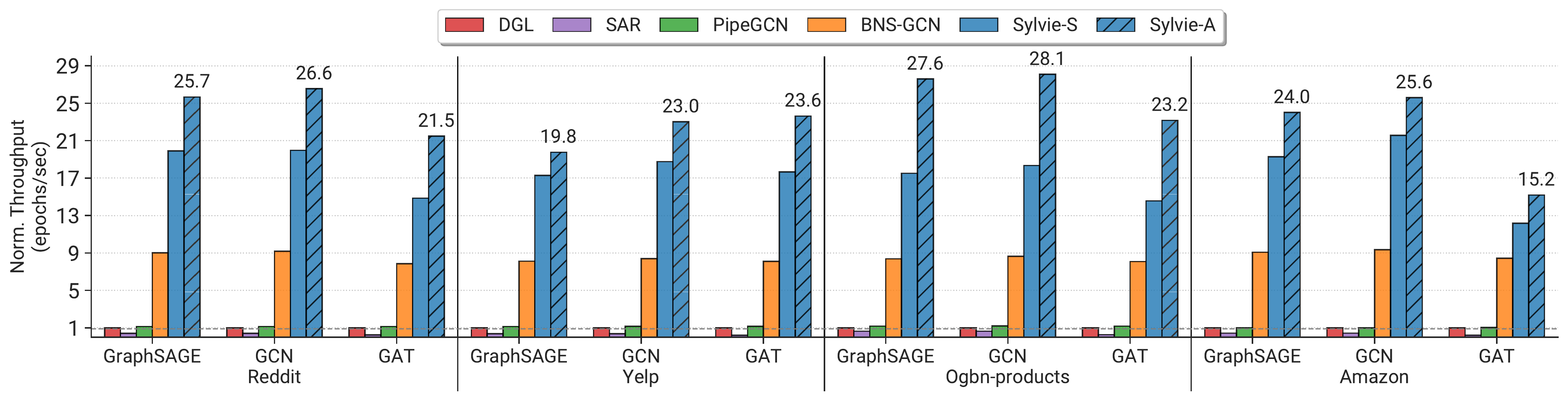}
    \vskip -10pt
    \caption{Training throughput of different methods (normalized to that of DGL, shown in the dashed line) when training three representative models on four datasets on two servers. \SysName outperforms DGL by up to 28.1$\times$.}
    \label{fig_multinode_speed}
    \vskip -10pt
\end{figure*}

\begin{table*}[t]
    \centering
    \caption{Detailed comparison of training throughput and test accuracy between \SysName and other baselines when training on two machines, where the best performance is highlighted in bold. \SysName always outperforms others in throughput on all the models and datasets.}
    \vskip 0.1in
    \resizebox{\linewidth}{!}{
        \begin{tabular}{ccrcrcrcrc}
            \toprule
            \multicolumn{1}{l}{}                & \multicolumn{1}{l}{} & \multicolumn{2}{c}{\textbf{Reddit}}                                & \multicolumn{2}{c}{\textbf{Yelp}}                                  & \multicolumn{2}{c}{\textbf{Ogbn-products}}                         & \multicolumn{2}{c}{\textbf{Amazon}}                                \\ \cmidrule(l){3-4}  \cmidrule(l){5-6} \cmidrule(l){7-8} \cmidrule(l){9-10}
            \textbf{Model}                      & \textbf{Method}      & \textbf{Thr.}          & \textbf{Test Acc.(\%)}                    & \textbf{Thr.}          & \textbf{F1-micro(\%)}                     & \textbf{Thr.}          & \textbf{Test Acc.(\%)}                    & \textbf{Thr.}          & \textbf{Test Acc.(\%)}                    \\ \midrule
            \multirow{6}{*}{\textbf{GraphSAGE}} & DGL                  & 1.00$\times$           & 97.10\textcolor{gray}{$\pm$0.01}          & 1.00$\times$           & 65.07\textcolor{gray}{$\pm$0.19}          & 1.00$\times$           & 79.19\textcolor{gray}{$\pm$0.15}          & 1.00$\times$           & \textbf{81.29}\textcolor{gray}{$\pm$0.02}          \\
                                                & SAR                  & 0.42$\times$           & 96.02\textcolor{gray}{$\pm$0.12}          & 0.37$\times$           & 60.51\textcolor{gray}{$\pm$0.09}          & 0.64$\times$           & 74.42\textcolor{gray}{$\pm$0.07}          & 0.43$\times$           & 78.85\textcolor{gray}{$\pm$0.07}          \\
                                                & PipeGCN              & 1.15$\times$           & 97.02\textcolor{gray}{$\pm$0.11}          & 1.15$\times$           & 65.14\textcolor{gray}{$\pm$0.08}          & 1.19$\times$           & \textbf{79.29}\textcolor{gray}{$\pm$0.05} & 1.05$\times$           & 81.27\textcolor{gray}{$\pm$0.08}          \\
                                                & BNS-GCN              & 9.02$\times$           & 97.14\textcolor{gray}{$\pm$0.01}          & 8.11$\times$           & \textbf{65.22}\textcolor{gray}{$\pm$0.23} & 8.38$\times$           & 79.11\textcolor{gray}{$\pm$0.11}          & 9.08$\times$           & 80.90\textcolor{gray}{$\pm$0.05}          \\
                                                & \SysNameS                & 19.90$\times$          & \textbf{97.15}\textcolor{gray}{$\pm$0.11} & 17.29$\times$          & 65.07\textcolor{gray}{$\pm$0.23}          & 17.51$\times$          & 78.86\textcolor{gray}{$\pm$0.17}          & 19.27$\times$          & 81.22\textcolor{gray}{$\pm$0.07}          \\
                                                & \SysNameP                & \textbf{25.66}$\times$ & 96.87\textcolor{gray}{$\pm$0.03}          & \textbf{19.76}$\times$ & 64.92\textcolor{gray}{$\pm$0.38}          & \textbf{27.59}$\times$ & 78.85\textcolor{gray}{$\pm$0.56}          & \textbf{24.01}$\times$ & 81.24\textcolor{gray}{$\pm$0.11} \\ \midrule
            \multirow{6}{*}{\textbf{GCN}}       & DGL                  & 1.00$\times$           & 94.84\textcolor{gray}{$\pm$0.58}          & 1.00$\times$           & 47.50\textcolor{gray}{$\pm$0.07}          & 1.00$\times$           & \textbf{74.58}\textcolor{gray}{$\pm$0.20}          & 1.00$\times$           & \textbf{56.59}\textcolor{gray}{$\pm$0.11}          \\
                                                & SAR                  & 0.42$\times$           & 95.34\textcolor{gray}{$\pm$0.17}          & 0.38$\times$           & 47.00\textcolor{gray}{$\pm$0.12}          & 0.65$\times$           & 70.13\textcolor{gray}{$\pm$0.10}          & 0.43$\times$           & 53.08\textcolor{gray}{$\pm$0.07}          \\
                                                & PipeGCN              & 1.15$\times$           & 94.69\textcolor{gray}{$\pm$0.56}          & 1.16$\times$           & 47.16\textcolor{gray}{$\pm$0.01}          & 1.20$\times$           & 74.04\textcolor{gray}{$\pm$0.23}          & 1.01$\times$           & 56.56\textcolor{gray}{$\pm$0.34}          \\
                                                & BNS-GCN              & 9.18$\times$           & 95.00\textcolor{gray}{$\pm$0.33}          & 8.40$\times$           & 47.27\textcolor{gray}{$\pm$0.37}          & 8.64$\times$           & 73.54\textcolor{gray}{$\pm$0.42}          & 9.34$\times$           & 56.47\textcolor{gray}{$\pm$0.60}          \\
                                                & \SysNameS                & 19.97$\times$          & \textbf{95.49}\textcolor{gray}{$\pm$0.04} & 18.76$\times$          & \textbf{48.77}\textcolor{gray}{$\pm$0.14} & 18.35$\times$          & 74.14\textcolor{gray}{$\pm$0.49} & 21.56$\times$          & 56.07\textcolor{gray}{$\pm$0.07} \\
                                                & \SysNameP                & \textbf{26.56}$\times$ & 95.31\textcolor{gray}{$\pm$0.01}          & \textbf{23.02}$\times$ & 47.62\textcolor{gray}{$\pm$0.30}          & \textbf{28.09}$\times$ & 73.78\textcolor{gray}{$\pm$0.19}          & \textbf{25.60}$\times$ & 55.84\textcolor{gray}{$\pm$0.21}          \\ \midrule
            \multirow{6}{*}{\textbf{GAT}}       & DGL                  & 1.00$\times$           & 93.97\textcolor{gray}{$\pm$0.60}          & 1.00$\times$           & 44.39\textcolor{gray}{$\pm$0.16}          & 1.00$\times$           & 78.14\textcolor{gray}{$\pm$0.12}          & 1.00$\times$           & 42.84\textcolor{gray}{$\pm$0.96}          \\
                                                & SAR                  & 0.25$\times$           & 91.47\textcolor{gray}{$\pm$0.08}          & 0.21$\times$           & 44.30\textcolor{gray}{$\pm$0.11}          & 0.27$\times$           & 76.40\textcolor{gray}{$\pm$0.06}          & 0.21$\times$           & 42.48\textcolor{gray}{$\pm$0.07}          \\
                                                & PipeGCN              & 1.14$\times$           & 94.49\textcolor{gray}{$\pm$0.64}          & 1.15$\times$           & 43.75\textcolor{gray}{$\pm$0.23}          & 1.19$\times$           & 77.03\textcolor{gray}{$\pm$0.11}          & 1.04$\times$           & 42.37\textcolor{gray}{$\pm$0.07}          \\
                                                & BNS-GCN              & 7.86$\times$           & 89.08\textcolor{gray}{$\pm$0.63}          & 8.11$\times$           & 43.66\textcolor{gray}{$\pm$0.24}          & 8.08$\times$           & 74.07\textcolor{gray}{$\pm$0.92}          & 8.43$\times$           & 40.67\textcolor{gray}{$\pm0.79$}          \\
                                                & \SysNameS                & 14.86$\times$          & \textbf{94.55}\textcolor{gray}{$\pm$0.77} & 17.66$\times$          & \textbf{44.44}\textcolor{gray}{$\pm$0.62} & 14.57$\times$          & 78.00\textcolor{gray}{$\pm$0.01}          & 12.18$\times$          & \textbf{42.95}\textcolor{gray}{$\pm$0.16} \\
                                                & \SysNameP                & \textbf{21.49}$\times$ & 93.40\textcolor{gray}{$\pm$0.62}          & \textbf{23.62}$\times$ & 43.15\textcolor{gray}{$\pm$0.63}          & \textbf{23.16}$\times$ & \textbf{78.38}\textcolor{gray}{$\pm$0.18} & \textbf{15.20}$\times$ & 41.83\textcolor{gray}{$\pm$0.25}          \\ \bottomrule
            \end{tabular}
            }
    \label{table_multinode}
    \vskip -3pt
\end{table*}

We first compare \SysName with other distributed full-graph GNN training methods in both the multi-node and single-node settings (\S\ref{eval_compare}). Then we present the convergence of \SysName on different datasets and models (\S\ref{eval_converge}). To explore how quantization affects the performance, we evaluate the accuracy and training time using different bit-widths for quantization (\S\ref{eval_bitwidth}). Finally, we analyze the overhead of \SysName (\S\ref{eval_overhead}). 

\textbf{Datasets and Models}. We evaluate \SysName on four real-world large-scale graph benchmarks: Reddit \cite{GraphSAGE}, Yelp \cite{GraphSAINT}, Ogbn-products \cite{OGB} and Amazon \cite{amazon}. We choose three popular GNN models, which are commonly adopted in evaluating GNN training: vanilla GCN \cite{GCN}, GraphSAGE \cite{GraphSAGE} and GAT \cite{GAT}.

\textbf{Baselines}. For the baselines, we compare \SysName with four SOTA distributed full-graph training methods: (1) DGL \cite{DGL}: the standard distributed GNN training on top of the latest DGL 0.9; (2) SAR \cite{SAR}; (3) PipeGCN \cite{PipeGCN}; (4) BNS-GCN \cite{BNS-GCN}: the $p$ value is set to $0.1$ as suggested by the paper. $p=0$ is not practical since it suffers from the worst test accuracy, the slowest convergence and severe overfitting. Baselines are orthogonal to each other in distributed GNN system designs so that we can make a fair comparison. 

\textbf{Testbeds}. We implement \SysName atop the latest stable version of popular GNN training library DGL 0.9 \cite{DGL} with PyTorch 1.10 \cite{PyTorch}. For all experiments, we use machines equipped with 8 GPUs (NVIDIA RTX 3090, each has 24GB GDDR6X Memory), one 32-thread CPU (AMD Threadripper PRO 3955WX) and 192GB DDR4 Memory. The intra-server communication (CPU-GPU and GPU-GPU) is based on PCIe 4.0 lanes. 



\subsection{Performance Evaluation} 
\label{eval_compare}
\textbf{Performance on Multiple Servers}. Training models on multiple machines is becoming a regular necessity nowadays. Fig.\ref{fig_multinode_speed} describes the throughput comparisons between \SysName and SOTA baselines on three models and four datasets over two machines. Here throughput is defined as the number of epochs run per second, and we normalize the throughput of each method on base of DGL. In each training task, we treat the first 10 epochs as the warmup stage and only record statistics afterward. We can clearly see that \SysName substantially outperforms other methods by a large margin on each dataset and model. Specifically, \SysNameS achieves a marvelous throughput improvement of 12.2$\sim$21.6$\times$ over DGL and far exceeds SAR and PipeGCN. Among the baselines, SAR shows the lowest throughput since it does not cope with the communication overhead, and its computation burden even increases due to the rematerialization. \SysNameS also delivers 1.4$\sim$2.3$\times$ larger throughput than BNS-GCN. 

\SysNameP further improves the training performance, which reaches 15.2$\sim$28.1$\times$ speedup over DGL and 1.8$\sim$3.3$\times$ speedup over BNS-GCN. In addition, we note that PipeGCN shows similar performance with DGL because in the multi-server training, the communication cost is immensely larger than computation. In this case, the communication could hardly be hidden so the performance gain is negligible. 

We show the detailed normalized training throughput and test accuracy in Table \ref{table_multinode}. In the multi-server setting, \SysName always achieves far better training throughput than other methods, demonstrating its effectiveness in large-scale distributed training. Additionally, the test accuracy of \SysName only suffers from minor degradation. In contrast, BNS-GCN incurs significant accuracy loss of up to 4.9\% compared to DGL on GAT, showing its limited generality to other models. The model accuracy of \SysNameP experiences a moderate fluctuation between $-$1.24\%$\sim$$+$1.25\%, due to the stale communicated embeddings \& gradients. 

\begin{figure}[t]
    \centering
    \includegraphics[width=\linewidth]{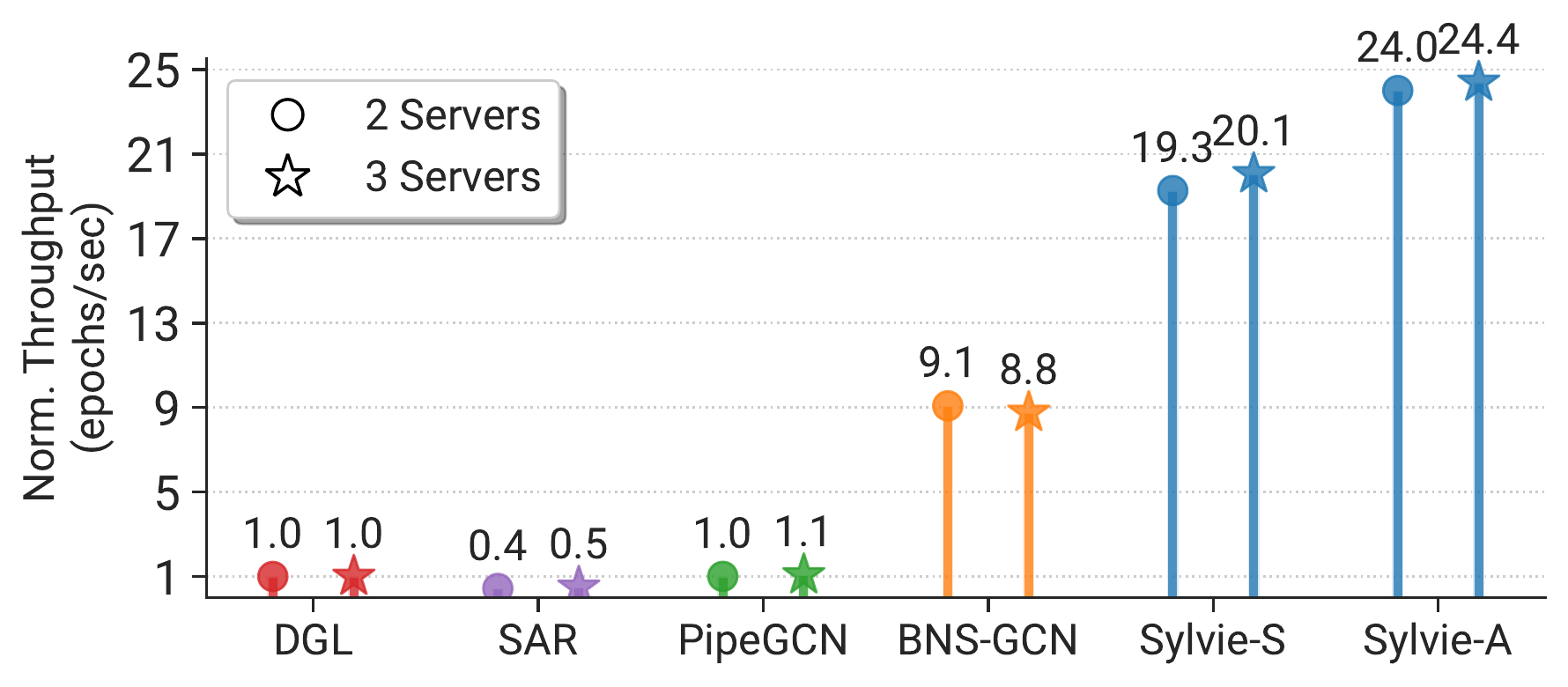}
    \vskip -10pt
    \caption{Normalized training throughput on two servers and three servers for GraphSAGE on Amazon.}
    \label{fig_multiserver_compare}
\end{figure}

Moreover, Fig.\ref{fig_multiserver_compare} presents the training throughput of \SysName in the two-server and three-server systems, respectively. We can observe that \SysName maintains the great performance and even achieves higher throughput acceleration ratio when the number of servers increases. On both settings, \SysName offers the best training speedup compared with other methods, while SAR and PipeGCN show very limited performance in large-scale training. In a nutshell, \SysName can deliver desired performance for larger-scale training scenarios.

\begin{figure}[t]
    \centering
    \includegraphics[width=\linewidth]{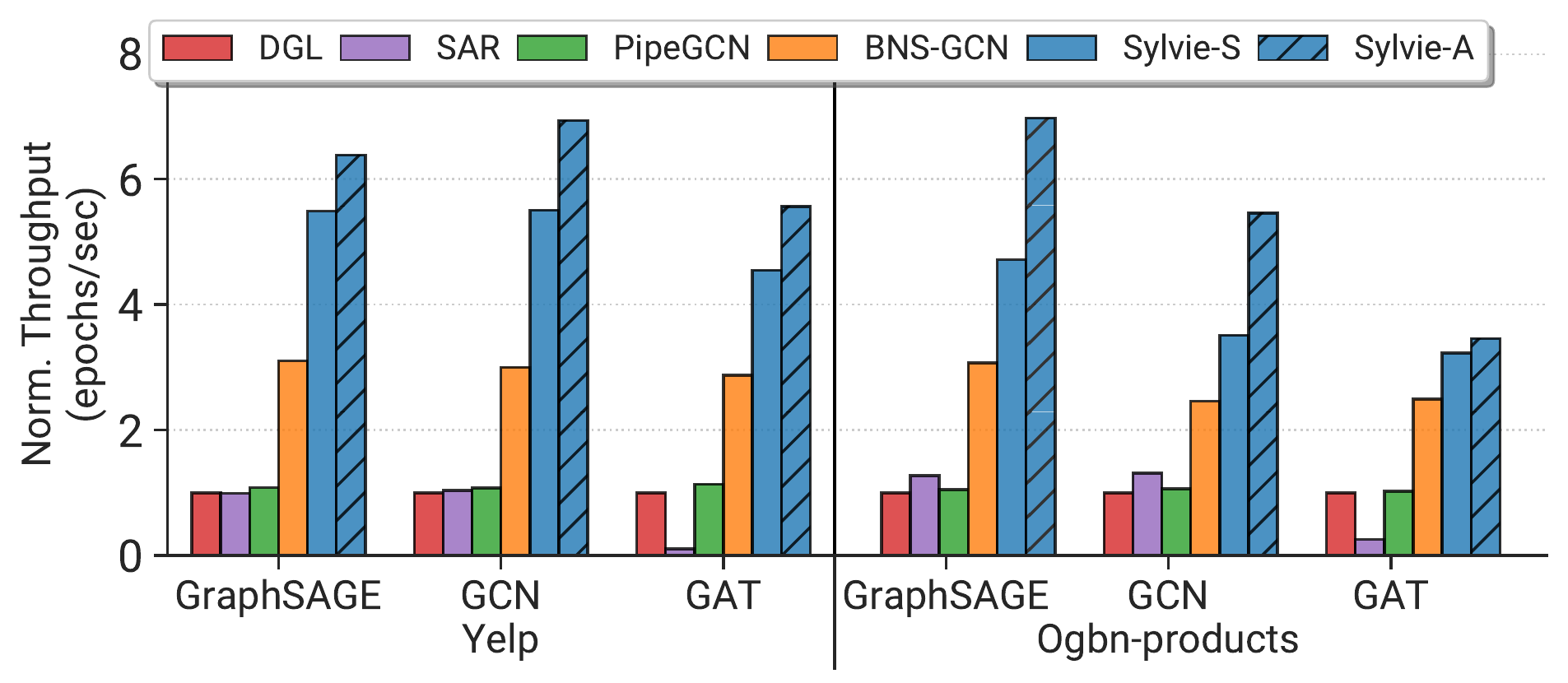}
    \vskip -10pt
    \caption{Some throughput results when training on a single machine with 8 GPUs. \SysName still outperforms others.}
    \label{fig_singlenode_speed}
\end{figure}

\textbf{Performance on Single Server}. We also test the performance of \SysName on one single machine with 8 GPUs. Fig.\ref{fig_singlenode_speed} shows partial results of the throughput for different methods due to the page limit. \SysName still outperforms other methods in training throughput, with a maximum of 7.0$\times$ speedup over DGL on Ogbn-products when training GraphSAGE. The speedup is relatively less significant compared to the multi-server setting, which involves more partitions with larger communication overhead. This indicates \SysName is more effective when the training scale is larger. 

\begin{table}[t]
    \centering
    \caption{Epoch communication volume and time breakdown of training GraphSAGE on four datasets over two servers. The communication volume decrease of \SysName is almost 32$\times$.}
    \vskip 0.1in
    \resizebox{\linewidth}{!}{
       \begin{tabular}{@{}cccccc@{}}
\toprule
\multicolumn{1}{l}{}                    & \multirow{2}{*}{\textbf{Method}} & \multicolumn{2}{c}{\textbf{Comm. Volume(MB)}}    &  \multicolumn{2}{c}{\textbf{Time per Epoch (s)}} \\ \cmidrule(l){3-4} \cmidrule(l){5-6}
                                        &                                  & \textbf{Main Data} & \textbf{Error-compensated} &  \textbf{Total} & \textbf{Comm.} \\ \midrule
\multirow{2}{*}{\textbf{Reddit}}        & DGL                              & 2791.7             & 0                          & 7.28                                              & 6.62                                             \\
                                        & \SysNameS                            & 87.3               & 10.7                       & 0.37                                              & 0.29                                             \\ \midrule
\multirow{2}{*}{\textbf{Yelp}}          & DGL                              & 2348.1             & 0                          & 4.98                                              & 4.70                                              \\
                                        & \SysNameS                            & 73.4               & 4.5                        & 0.29                                              & 0.21                                             \\ \midrule
\multirow{2}{*}{\textbf{Ogbn-products}} & DGL                              & 3420.6             & 0                          & 6.03                                              & 5.87                                             \\
                                        & \SysNameS                            & 106.9              & 26.6                       & 0.34                                              & 0.22                                             \\ \midrule
\multirow{2}{*}{\textbf{Amazon}}        & DGL                              & 5632.6             & 0                          & 13.33                                             & 11.47                                            \\
                                        & \SysNameS                            & 176.1              & 11                         & 0.69                                              & 0.57                                             \\ \bottomrule
\end{tabular}
            }
    \label{table_commu_volume}
    \vskip 7pt
\end{table}

\textbf{Communication Volume and Time}. To demonstrate the training speedup is due to the reduced communication, we record the actual communication volume per epoch and training time breakdown in Table \ref{table_commu_volume}. We observe that \SysName cuts down the communication volume dramatically. For example, there are originally 5632.6 MB communication per epoch for the Amazon dataset. After deploying \SysName, there are only 176.1 MB communicated embeddings \& gradients, reducing almost 32$\times$ communication volume. Accordingly, the communication time is vastly shortened. The time breakdown shows the communication occupies a very large portion in epoch time, so the training throughput is also improved. Note that we also transmit some error-compensated information, which accounts for much smaller portions (e.g., 11.0 MB for Amazon) and incurs negligible overhead.

\subsection{Impact on Convergence}
\label{eval_converge}

\begin{figure}[t]
    \centering
    \includegraphics[width=\linewidth]{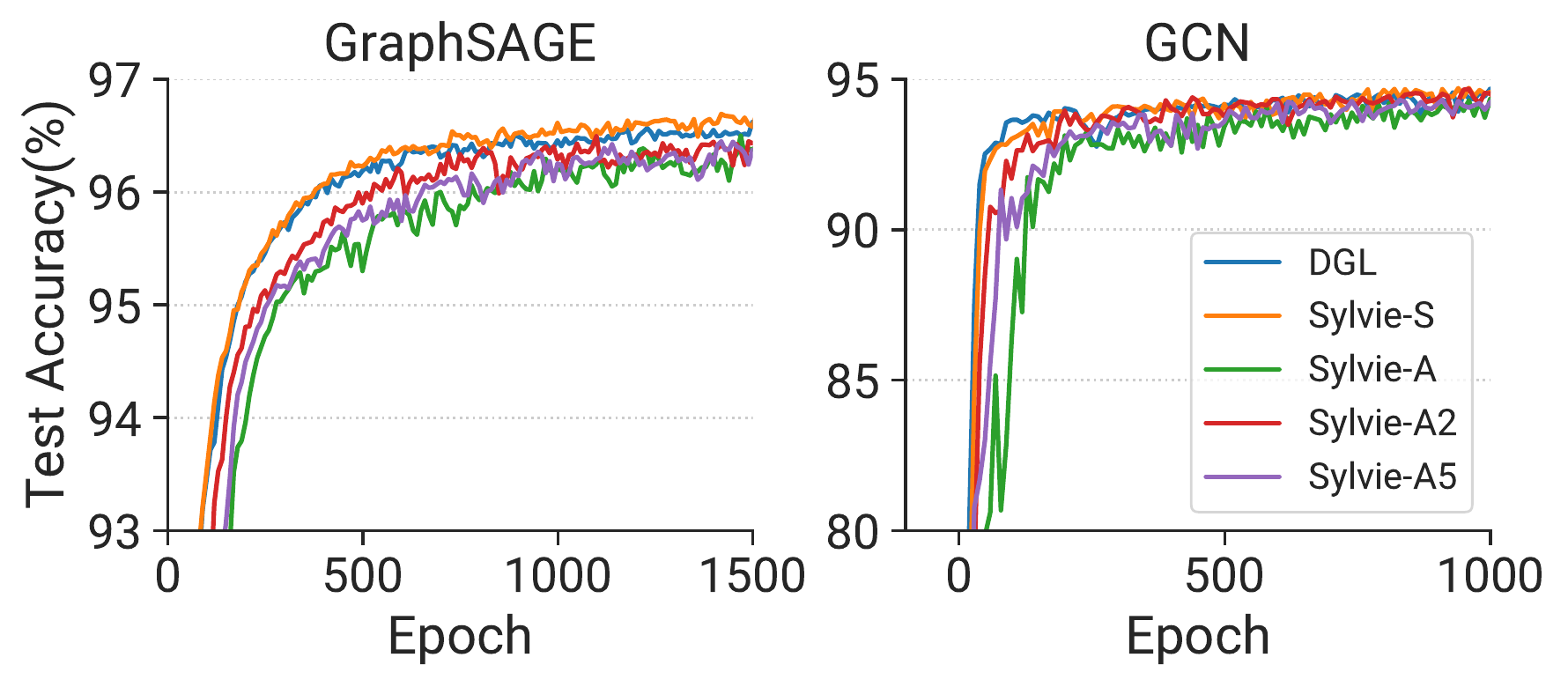}
    \vskip -10pt
    \caption{Test accuracy convergence during the training process with DGL, \SysNameS, \SysNameP and \SysNameP with \textit{Bounded Staleness Adaptor} ($\epsilon_s=\{2,5\}$) on Reddit.}
    \label{fig_stale_adaptor}
\end{figure}

We examine the convergence curves of \SysName in Fig.\ref{fig_stale_adaptor}. We can see the curves of \SysNameS are almost identical to that of DGL in this case. However, \SysNameP gives slower convergence rate at an early phase of some datasets or models. To mitigate the errors by stale embeddings \& feature gradients, we train \SysNameP with \emph{Bounded Staleness Adaptor} when $\epsilon_s=\{2,5\}$ on Reddit. We observe that with periodical synchronization of the latest data, \SysNameP boosts the convergence speed well, and the convergence curve approximates that of vanilla training more when the synchronization is more frequent ($\epsilon_s$ gets smaller).


\begin{figure}[t]
    \centering
    \includegraphics[width=0.9\linewidth]{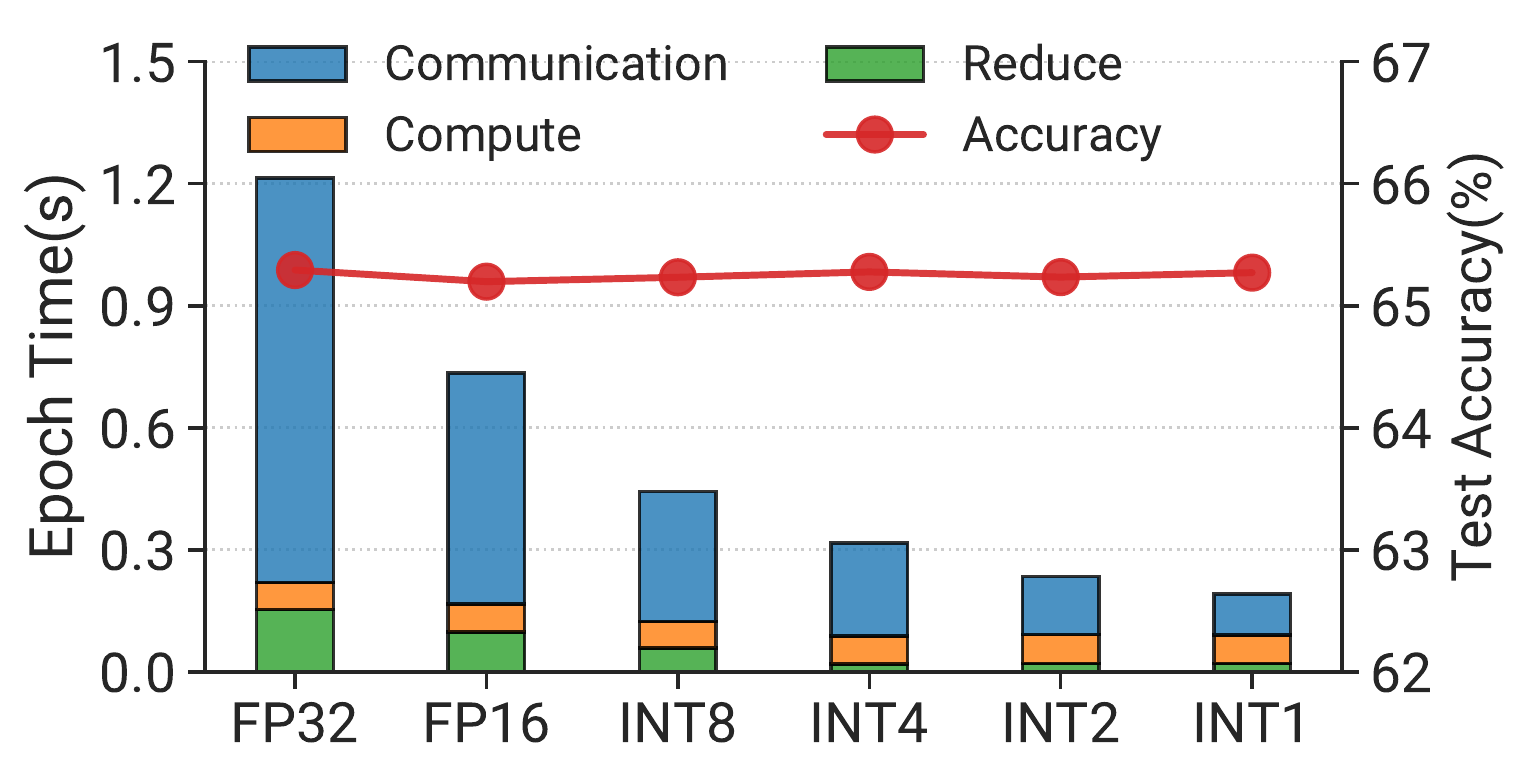}
    \vskip -10pt
    \caption{Training time per epoch, its breakdown and corresponding test accuracy when using different quantization bit-widths to train GraphSAGE on Yelp over single server (N=8).}
    \label{fig_quant_bit}
    \vskip -5pt
\end{figure}

\subsection{Quantization Effect Analysis}
\label{eval_bitwidth}
\textbf{Impact of the Bit-width}. To explore how the quantization bit-width $b$ contributes to the training throughput improvement, we showcase the training epoch time along with its breakdown for GraphSAGE on \SysNameS with different quantization bit-widths in Fig.\ref{fig_quant_bit}. As the bit-width goes down, there occurs semi-linear reduction in the communication overhead, thus leading to a speedup on the epoch time. The vanilla method uses FP32 datatype for communication, which occupies nearly the whole epoch time (0.99s out of 1.21s). Switching to FP16 datatype, the communication cost almost decreases by half. When using 1-bit quantization, we cut down almost 89.8\% communication overhead and 84.2\% training time per epoch compared to the vanilla method. 

\textbf{Impact of the Quantization Ratio}.To further prove the effectiveness of our method on the model performance, we conduct training by quantizing \textbf{all} the embeddings \& feature gradients to 1-bit. Table \ref{table_quantall} shows the comparisons of test accuracy between \SysNameS and quantizing all data. On all datasets and models, quantizing all data will bring serious accuracy loss, e.g. 97.2\% in \SysNameS versus 70.6\% in quantizing all. The accuracy drop is due to the overmuch distortion on data by absolute quantization, deteriorating both the forward and backward pass in training. 

\begin{table}[t]
    \centering
    \caption{Test accuracy comparison when training with \SysName and quantizing all embeddings \& feature gradients.}
    \vskip 0.1in
    \resizebox{\linewidth}{!}{
      \begin{tabular}{ccccccc}
\toprule
                   & \multicolumn{2}{c}{\textbf{Reddit}} & \multicolumn{2}{c}{\textbf{Yelp}} & \multicolumn{2}{c}{\textbf{Ogbn-products}} \\ \cmidrule(l){2-3} \cmidrule(l){4-5} \cmidrule(l){6-7} 
                   & \textbf{\SysNameS}  & \textbf{Quant All}  & \textbf{\SysNameS} & \textbf{Quant All} & \textbf{\SysNameS}      & \textbf{Quant All}     \\ \midrule
\textbf{GraphSAGE} & 97.15         & 70.63               & 65.07        & 57.14              & 78.86             & 58.25                  \\
\textbf{GCN}       & 95.49         & 92.60                & 48.77        & 42.24              & 74.14             & 61.25                  \\
\textbf{GAT}       & 94.55         & 91.72               & 44.44        & 25.76              & 78.00             & 69.95                  \\ \bottomrule
\end{tabular}
            }
    \label{table_quantall}
\end{table}

\subsection{Overhead Analysis}
\label{eval_overhead}
\begin{figure}[t]
    \centering
    \includegraphics[width=\linewidth]{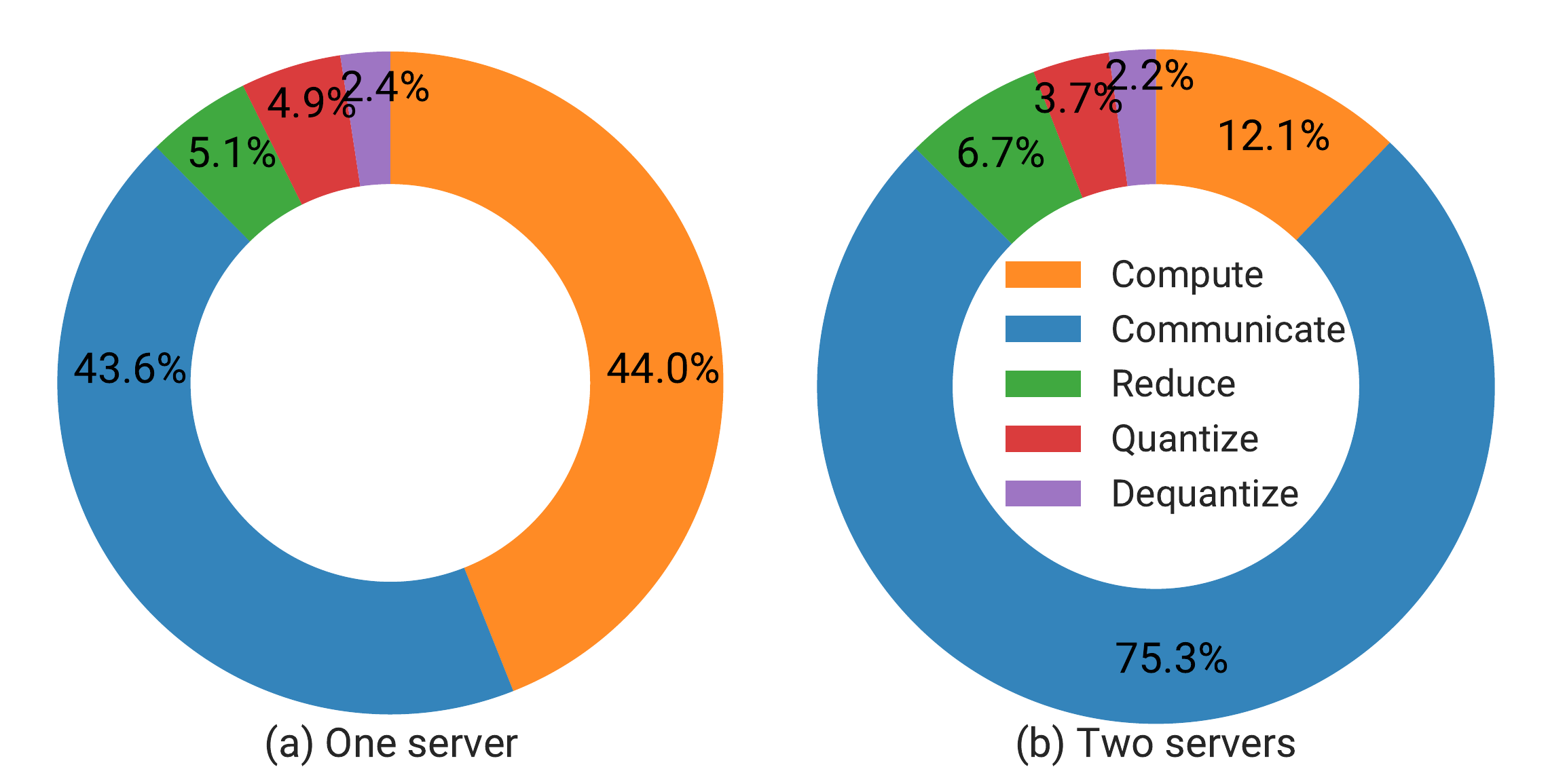}
    \vskip -10pt
    \caption{Ratios of different components in epoch time when training GraphSAGE with \SysNameS on Reddit over single server and two servers. \textit{Low-bit Module} has negligible overhead.}
    \label{fig_piechart}
\end{figure}

\textbf{Quantization Overhead}. To explore how much overhead the \textit{Low-bit module} takes up and find the latent optimization opportunities, we record the time spent on each part in an epoch with \SysNameS on single server and two servers in Fig.\ref{fig_piechart}. Both cases demonstrate that the time consumed by this module occupies the smallest portions, indicating the negligible overhead brought by our methods. Specifically, under the two-server setting, the total ratio of them (5.9\%) are even smaller than that of the all-reduce (6.7\%). And the communication overhead becomes more dominant in the multi-server setting.

%% file: 5_Discussion.tex
\section{Future Work}
\label{futurework}

\textbf{Error Compensation}. 
Inevitably, quantization and dequantization process will introduce errors into the training process. These errors of embeddings and embedding gradients will even accumulate in the subsequent layers, seriously affecting the model quality when GNN model becomes deep. In our future work, we can utilize two observations to adjust the trade-off between accuracy and training throughput. The first observation is early training epochs can use lower bit-width quantization while latter rounds with higher bit-width without affecting convergence. In this way, we can automatically adjust the bit-width according to the status of training process. Another observation comes from the distribution property of nodes. In GNN layers, the aggregation phase is the source of substantial error, especially at nodes with higher degrees. As the degree of nodes increases, the variance of aggregation values will also increase. Therefore, we can consider applying different bit-width quantization in the future to different nodes according to their importance. 

\textbf{Memory Footprint}. Besides offering substantial improvements on the training throughput, the possibility of reducing the memory footprint by quantization is also worth exploring. Using low-precision values is expected to reduce the maximum memory allocated. For instance, when training GraphSAGE on the Ogbn-products dataset, the maximum memory used is about 6882 MB for full-precision FP32 communication and 6642 MB for half-precision FP16 communication. However, because we conduct 1-bit quantization and dequantization using GPUs, we find there are some extra memory expenses during training. The maximum memory used by \SysNameS reaches 6918 MB, which is slightly higher than vanilla training. Nevertheless, using quantization to reduce memory footprints is still a promising direction and we leave this as a potential future work. 

%% file: 7_Conclusion.tex
\section{Conclusion}
\label{sec_conclusion}
This work proposes \SysName, an efficient distributed GNN training framework that enormously reduces the communication cost by quantizing the communicated data to low bit-width values while maintaining the model quality as much as possible. We also integrate \SysName with the asynchronous pipeline technique and \textit{Bounded Staleness Adaptor} to further enhance the training performance. Extensive experiments show that \SysName can substantially boost the training throughput by up to 28.1$\times$.


%% file: main.bbl
\begin{thebibliography}{42}
\providecommand{\natexlab}[1]{#1}
\providecommand{\url}[1]{\texttt{#1}}
\expandafter\ifx\csname urlstyle\endcsname\relax
  \providecommand{\doi}[1]{doi: #1}\else
  \providecommand{\doi}{doi: \begingroup \urlstyle{rm}\Url}\fi

\bibitem[Lip(2022)]{LipschitzContinuity}
Lipschitz continuity.
\newblock \url{https://en.wikipedia.org/wiki/Lipschitz_continuity}, 2022.

\bibitem[cuS(2022)]{cuSPARSE}
Nvidia cuda sparse matrix library.
\newblock \url{https://docs.nvidia.com/cuda/cusparse/index.html#abstract},
  2022.

\bibitem[Abadal et~al.(2021)Abadal, Jain, Guirado, López-Alonso, and
  Alarcón]{GNN-survey}
Abadal, S., Jain, A., Guirado, R., López-Alonso, J., and Alarcón, E.
\newblock Computing graph neural networks: A survey from algorithms to
  accelerators.
\newblock \emph{CoRR}, abs/2010.00130, 2021.

\bibitem[Alistarh et~al.(2017)Alistarh, Grubic, Li, Tomioka, and
  Vojnovic]{QSGD}
Alistarh, D., Grubic, D., Li, J., Tomioka, R., and Vojnovic, M.
\newblock Qsgd: Communication-efficient sgd via gradient quantization and
  encoding.
\newblock In \emph{Advances in Neural Information Processing Systems}, NeurIPS
  '17, 2017.

\bibitem[Bottou et~al.(2018)Bottou, Curtis, and
  Nocedal]{bottou2018optimization}
Bottou, L., Curtis, F.~E., and Nocedal, J.
\newblock Optimization methods for large-scale machine learning.
\newblock \emph{CoRR}, abs/1606.04838, 2018.

\bibitem[Chen et~al.(2018)Chen, Ma, and Xiao]{FastGCN}
Chen, J., Ma, T., and Xiao, C.
\newblock Fastgcn: Fast learning with graph convolutional networks via
  importance sampling.
\newblock In \emph{International Conference on Learning Representations}, ICLR
  '18, 2018.

\bibitem[Chen et~al.(2021)Chen, Zheng, Yao, Wang, Stoica, Mahoney, and
  Gonzalez]{ActNN}
Chen, J., Zheng, L., Yao, Z., Wang, D., Stoica, I., Mahoney, M., and Gonzalez,
  J.
\newblock Actnn: Reducing training memory footprint via 2-bit activation
  compressed training.
\newblock In \emph{Proceedings of the 38th International Conference on Machine
  Learning}, ICML '21, 2021.

\bibitem[Chiang et~al.(2019)Chiang, Liu, Si, Li, Bengio, and Hsieh]{ClusterGCN}
Chiang, W.-L., Liu, X., Si, S., Li, Y., Bengio, S., and Hsieh, C.-J.
\newblock Cluster-gcn: An efficient algorithm for training deep and large graph
  convolutional networks.
\newblock In \emph{Proceedings of the 25th ACM SIGKDD International Conference
  on Knowledge Discovery \& Data Mining}, KDD '19, 2019.

\bibitem[Courbariaux et~al.(2015)Courbariaux, Bengio, and David]{BinaryConnect}
Courbariaux, M., Bengio, Y., and David, J.-P.
\newblock Binaryconnect: Training deep neural networks with binary weights
  during propagations.
\newblock In \emph{Advances in Neural Information Processing Systems}, NeurIPS
  '15, 2015.

\bibitem[Dong et~al.(2017)Dong, Ni, Li, Chen, Zhu, and Su]{SQ}
Dong, Y., Ni, R., Li, J., Chen, Y., Zhu, J., and Su, H.
\newblock Learning accurate low-bit deep neural networks with stochastic
  quantization.
\newblock \emph{CoRR}, abs/1708.01001, 2017.

\bibitem[Feng et~al.(2020)Feng, Wang, Li, Yang, Peng, and Ding]{SGQuant}
Feng, B., Wang, Y., Li, X., Yang, S., Peng, X., and Ding, Y.
\newblock Sgquant: Squeezing the last bit on graph neural networks with
  specialized quantization.
\newblock In \emph{2020 IEEE 32nd International Conference on Tools with
  Artificial Intelligence (ICTAI)}, 2020.

\bibitem[Fey \& Lenssen(2019)Fey and Lenssen]{PyG}
Fey, M. and Lenssen, J.~E.
\newblock Fast graph representation learning with pytorch geometric.
\newblock \emph{CoRR}, abs/1903.02428, 2019.

\bibitem[Hamilton et~al.(2017)Hamilton, Ying, and Leskovec]{GraphSAGE}
Hamilton, W., Ying, Z., and Leskovec, J.
\newblock Inductive representation learning on large graphs.
\newblock In \emph{Advances in Neural Information Processing Systems}, NeurIPS
  '17, 2017.

\bibitem[He \& McAuley(2016)He and McAuley]{amazon}
He, R. and McAuley, J.
\newblock Ups and downs: Modeling the visual evolution of fashion trends with
  one-class collaborative filtering.
\newblock In \emph{Proceedings of the 25th International Conference on World
  Wide Web}, WWW '16, 2016.

\bibitem[Hu et~al.(2020)Hu, Fey, Zitnik, Dong, Ren, Liu, Catasta, and
  Leskovec]{OGB}
Hu, W., Fey, M., Zitnik, M., Dong, Y., Ren, H., Liu, B., Catasta, M., and
  Leskovec, J.
\newblock Open graph benchmark: Datasets for machine learning on graphs.
\newblock In \emph{Advances in Neural Information Processing Systems}, NeurIPS
  '20, 2020.

\bibitem[Huang et~al.(2018)Huang, Zhang, Rong, and Huang]{Adaptive18}
Huang, W., Zhang, T., Rong, Y., and Huang, J.
\newblock Adaptive sampling towards fast graph representation learning.
\newblock In \emph{Proceedings of the 32nd International Conference on Neural
  Information Processing Systems}, NIPS'18, 2018.

\bibitem[Jacob et~al.(2018)Jacob, Kligys, Chen, Zhu, Tang, Howard, Adam, and
  Kalenichenko]{QuantizeInfer}
Jacob, B., Kligys, S., Chen, B., Zhu, M., Tang, M., Howard, A., Adam, H., and
  Kalenichenko, D.
\newblock Quantization and training of neural networks for efficient
  integer-arithmetic-only inference.
\newblock In \emph{Proceedings of the IEEE Conference on Computer Vision and
  Pattern Recognition}, CVPR '18, 2018.

\bibitem[Jia et~al.(2020)Jia, Lin, Gao, Zaharia, and Aiken]{ROC}
Jia, Z., Lin, S., Gao, M., Zaharia, M., and Aiken, A.
\newblock Improving the accuracy, scalability, and performance of graph neural
  networks with roc.
\newblock In \emph{Proceedings of Machine Learning and Systems}, MLSys '20,
  2020.

\bibitem[Kipf \& Welling(2016)Kipf and Welling]{GCN}
Kipf, T.~N. and Welling, M.
\newblock Semi-supervised classification with graph convolutional networks.
\newblock In \emph{International Conference on Learning Representations}, ICLR
  '16, 2016.

\bibitem[Krishnamoorthi(2018)]{QuantizeDNN}
Krishnamoorthi, R.
\newblock Quantizing deep convolutional networks for efficient inference: A
  whitepaper.
\newblock \emph{CoRR}, abs/1806.08342, 2018.

\bibitem[Liu et~al.(2021)Liu, Zhou, Yang, Li, Chen, and Hu]{EXACT}
Liu, Z., Zhou, K., Yang, F., Li, L., Chen, R., and Hu, X.
\newblock Exact: Scalable graph neural networks training via extreme activation
  compression.
\newblock In \emph{International Conference on Learning Representations}, ICLR
  '21, 2021.

\bibitem[Ma et~al.(2019)Ma, Yang, Miao, Xue, Wu, Zhou, and Dai]{NeuGraph}
Ma, L., Yang, Z., Miao, Y., Xue, J., Wu, M., Zhou, L., and Dai, Y.
\newblock {NeuGraph}: Parallel deep neural network computation on large graphs.
\newblock In \emph{2019 USENIX Annual Technical Conference}, {USENIX} {ATC}
  '19, 2019.

\bibitem[Mostafa(2022)]{SAR}
Mostafa, H.
\newblock Sequential aggregation and rematerialization: Distributed full-batch
  training of graph neural networks on large graphs.
\newblock In \emph{Proceedings of Machine Learning and Systems}, MLSys '22,
  2022.

\bibitem[Paszke et~al.(2019)Paszke, Gross, Massa, Lerer, Bradbury, Chanan,
  Killeen, Lin, Gimelshein, Antiga, Desmaison, Kopf, Yang, DeVito, Raison,
  Tejani, Chilamkurthy, Steiner, Fang, Bai, and Chintala]{PyTorch}
Paszke, A., Gross, S., Massa, F., Lerer, A., Bradbury, J., Chanan, G., Killeen,
  T., Lin, Z., Gimelshein, N., Antiga, L., Desmaison, A., Kopf, A., Yang, E.,
  DeVito, Z., Raison, M., Tejani, A., Chilamkurthy, S., Steiner, B., Fang, L.,
  Bai, J., and Chintala, S.
\newblock Pytorch: An imperative style, high-performance deep learning library.
\newblock In \emph{Advances in Neural Information Processing Systems}, NeurIPS
  '19, 2019.

\bibitem[Ramezani et~al.(2022)Ramezani, Cong, Mahdavi, Kandemir, and
  Sivasubramaniam]{LLCG}
Ramezani, M., Cong, W., Mahdavi, M., Kandemir, M., and Sivasubramaniam, A.
\newblock Learn locally, correct globally: A distributed algorithm for training
  graph neural networks.
\newblock In \emph{International Conference on Learning Representations}, ICLR
  '22, 2022.

\bibitem[Stock et~al.(2020)Stock, Fan, Graham, Grave, Gribonval, Jegou, and
  Joulin]{QuantNoise}
Stock, P., Fan, A., Graham, B., Grave, E., Gribonval, R., Jegou, H., and
  Joulin, A.
\newblock Training with quantization noise for extreme model compression.
\newblock In \emph{International Conference on Learning Representations}, ICLR
  '20, 2020.

\bibitem[Tailor et~al.(2021)Tailor, Fernandez-Marques, and Lane]{Degree-Quant}
Tailor, S.~A., Fernandez-Marques, J., and Lane, N.~D.
\newblock Degree-quant: Quantization-aware training for graph neural networks.
\newblock \emph{CoRR}, abs/2008.05000, 2021.

\bibitem[Thorpe et~al.(2021)Thorpe, Qiao, Eyolfson, Teng, Hu, Jia, Wei, Vora,
  Netravali, Kim, and Xu]{Dorylus}
Thorpe, J., Qiao, Y., Eyolfson, J., Teng, S., Hu, G., Jia, Z., Wei, J., Vora,
  K., Netravali, R., Kim, M., and Xu, G.~H.
\newblock Dorylus: Affordable, scalable, and accurate {GNN} training with
  distributed {CPU} servers and serverless threads.
\newblock In \emph{15th USENIX Symposium on Operating Systems Design and
  Implementation}, OSDI '21, 2021.

\bibitem[Tripathy et~al.(2020)Tripathy, Yelick, and Buluç]{CAGNET}
Tripathy, A., Yelick, K., and Buluç, A.
\newblock Reducing communication in graph neural network training.
\newblock In \emph{SC20: International Conference for High Performance
  Computing, Networking, Storage and Analysis}, 2020.

\bibitem[Veličković et~al.(2018)Veličković, Cucurull, Casanova, Romero,
  Liò, and Bengio]{GAT}
Veličković, P., Cucurull, G., Casanova, A., Romero, A., Liò, P., and Bengio,
  Y.
\newblock Graph attention networks.
\newblock In \emph{International Conference on Learning Representations}, ICLR
  '18, 2018.

\bibitem[Wan et~al.(2022{\natexlab{a}})Wan, Li, Li, Kim, and Lin]{BNS-GCN}
Wan, C., Li, Y., Li, A., Kim, N.~S., and Lin, Y.
\newblock Bns-gcn: Efficient full-graph training of graph convolutional
  networks with partition-parallelism and random boundary node sampling.
\newblock In \emph{Proceedings of Machine Learning and Systems}, MLSys '22,
  2022{\natexlab{a}}.

\bibitem[Wan et~al.(2022{\natexlab{b}})Wan, Li, Wolfe, Kyrillidis, Kim, and
  Lin]{PipeGCN}
Wan, C., Li, Y., Wolfe, C.~R., Kyrillidis, A., Kim, N.~S., and Lin, Y.
\newblock Pipegcn: Efficient full-graph training of graph convolutional
  networks with pipelined feature communication.
\newblock In \emph{International Conference on Learning Representations}, ICLR
  '22, 2022{\natexlab{b}}.

\bibitem[Wang et~al.(2020)Wang, Zheng, Ye, Gan, Li, Song, Zhou, Ma, Yu, Gai,
  Xiao, He, Karypis, Li, and Zhang]{DGL}
Wang, M., Zheng, D., Ye, Z., Gan, Q., Li, M., Song, X., Zhou, J., Ma, C., Yu,
  L., Gai, Y., Xiao, T., He, T., Karypis, G., Li, J., and Zhang, Z.
\newblock Deep graph library: A graph-centric, highly-performant package for
  graph neural networks.
\newblock \emph{CoRR}, abs/1909.01315, 2020.

\bibitem[Wang et~al.(2022)Wang, Feng, and Ding]{QGTC}
Wang, Y., Feng, B., and Ding, Y.
\newblock Qgtc: accelerating quantized graph neural networks via gpu tensor
  core.
\newblock In \emph{Proceedings of the 27th ACM SIGPLAN Symposium on Principles
  and Practice of Parallel Programming}, PPoPP '22, 2022.

\bibitem[Wen et~al.(2017)Wen, Xu, Yan, Wu, Wang, Chen, and Li]{TernGrad}
Wen, W., Xu, C., Yan, F., Wu, C., Wang, Y., Chen, Y., and Li, H.
\newblock Terngrad: Ternary gradients to reduce communication in distributed
  deep learning.
\newblock In \emph{Advances in Neural Information Processing Systems}, NeurIPS
  '17, 2017.

\bibitem[Xian et~al.(2021)Xian, Li, Liu, Guo, and Du]{H-PS}
Xian, L., Li, B., Liu, J., Guo, Z., and Du, D. H.~C.
\newblock H-ps: A heterogeneous-aware parameter server with distributed neural
  network training.
\newblock \emph{IEEE Access}, 9:\penalty0 44049--44058, 2021.

\bibitem[Zeng et~al.(2020)Zeng, Zhou, Srivastava, Kannan, and
  Prasanna]{GraphSAINT}
Zeng, H., Zhou, H., Srivastava, A., Kannan, R., and Prasanna, V.
\newblock Graphsaint: Graph sampling based inductive learning method.
\newblock In \emph{International Conference on Learning Representations}, ICLR
  '20, 2020.

\bibitem[Zhang \& Chen(2018)Zhang and Chen]{Link18}
Zhang, M. and Chen, Y.
\newblock Link prediction based on graph neural networks.
\newblock In \emph{Advances in Neural Information Processing Systems}, NeurIPS
  '18, 2018.

\bibitem[Zhao et~al.(2020)Zhao, Wang, Bates, Mullins, Jamnik, and Lio]{LowGNN}
Zhao, Y., Wang, D., Bates, D., Mullins, R., Jamnik, M., and Lio, P.
\newblock Learned low precision graph neural networks.
\newblock \emph{CoRR}, abs/2009.09232, 2020.

\bibitem[Zheng et~al.(2021)Zheng, Ma, Wang, Zhou, Su, Song, Gan, Zhang, and
  Karypis]{DistDGL}
Zheng, D., Ma, C., Wang, M., Zhou, J., Su, Q., Song, X., Gan, Q., Zhang, Z.,
  and Karypis, G.
\newblock Distdgl: Distributed graph neural network training for billion-scale
  graphs.
\newblock \emph{CoRR}, abs/2010.05337, 2021.

\bibitem[Zhu et~al.(2019)Zhu, Zhao, Yang, Lin, Zhou, Ai, Li, and
  Zhou]{AliGraph}
Zhu, R., Zhao, K., Yang, H., Lin, W., Zhou, C., Ai, B., Li, Y., and Zhou, J.
\newblock Aligraph: a comprehensive graph neural network platform.
\newblock \emph{Proceedings of the VLDB Endowment}, 12:\penalty0 2094--2105,
  2019.

\bibitem[Zou et~al.(2019)Zou, Hu, Wang, Jiang, Sun, and Gu]{Layerdependent19}
Zou, D., Hu, Z., Wang, Y., Jiang, S., Sun, Y., and Gu, Q.
\newblock Layer-dependent importance sampling for training deep and large graph
  convolutional networks.
\newblock In \emph{Proceedings of the 33rd International Conference on Neural
  Information Processing Systems}, NIPS'19, 2019.

\end{thebibliography}
